\documentclass[journal]{IEEEtran}
\usepackage{subfigure}
\usepackage{color}
\usepackage{cite}
\usepackage{amsmath}
\usepackage{cite}
\usepackage{graphicx}
\usepackage{epstopdf}
\usepackage{amsfonts,amsmath,amssymb}
\usepackage{graphicx}
\usepackage{url}
\usepackage{bm}
\usepackage{bbm}
\usepackage{subfigure}
\usepackage{stfloats}
\usepackage{color}

\usepackage{cite}

\newcommand{\Hnull}{\mathcal{H}_0}
\newcommand{\Halt}{\mathcal{H}_1}

\newcommand{\Honull}{\mathcal{{D}}_0}
\newcommand{\Hoalt}{\mathcal{{D}}_1}

\usepackage{cite,graphicx,amsmath,amssymb,cite,algorithm}

\newtheorem{corollary}{\textbf{Corollary}}
\newtheorem{theorem}{\textbf{Theorem}}

\newtheorem{lemma}{\textbf{Lemma}}
\newtheorem{remark}{\textbf{Remark}}

\begin{document}

\title{Covert Transmission with a Self-sustained Relay}
\author{{Jinsong Hu, Shihao Yan, Feng Shu, and Jiangzhou Wang}

\thanks{This work was supported in part by the National Natural Science Foundation of China under Grant 61771244 and Grant 61472190.
}

\thanks{J. Hu is with the College of Physics and Information, Fuzhou University, Fuzhou 350116, China, and also with the School of Electronic and Optical Engineering, Nanjing University of Science and Technology, Nanjing, Jiangsu 210094, China (e-mail: jinsong.hu@fzu.edu.cn). }

\thanks{S. Yan is with the School of Engineering, Macquarie University, Sydney, NSW 2109, Australia (e-mail: shihao.yan@mq.edu.au).}

\thanks{F. Shu is with the School of Electronic and Optical Engineering, Nanjing University of Science and Technology, Nanjing, Jiangsu 210094, China, and also with the College of Physics and Information, Fuzhou University, Fuzhou 350116, China (e-mail: shufeng@njust.edu.cn). }

\thanks{J. Wang is with the School of Engineering and Digital Arts, University of Kent, Canterbury CT2 7NT, U.K. (e-mail: j.z.wang@kent.ac.uk).}

}
\maketitle

\vspace{-2cm}

\begin{abstract}
This work examines the possibility, performance limits, and associated costs for a self-sustained relay to transmit its own covert information to a destination on top of forwarding the source's information. Since the source provides energy to the relay for forwarding its information, the source does not allow the relay's covert transmission and is to detect it. Considering the time switching (TS) and power splitting (PS) schemes for energy harvesting, where all the harvested energy is used for transmission at the self-sustained relay, we derive the minimum detection error probability $\xi^{\ast}$ at the source, based on which we determine the maximum effective covert rate $\Psi^{\ast}$ subject to a given covertness constraint on $\xi^{\ast}$. Our analysis shows that $\xi^{\ast}$ is the same for the TS and PS schemes, which leads to the fact that the cost of achieving $\Psi^{\ast}$ in both the two schemes in terms of the required increase in the energy conversion efficiency at the relay is the same, although the values of $\Psi^{\ast}$ in these two schemes can be different in specific scenarios. For example, the TS scheme outperforms the PS scheme in terms of achieving a higher $\Psi^{\ast}$ when the transmit power at the source is relatively low. If the covertness constraint is tighter than a specific value, it is the covertness constraint that limits $\Psi^{\ast}$, and otherwise it is upper bound on the energy conversion efficiency that limits $\Psi^{\ast}$.
\end{abstract}

%
%
\section{Introduction}

Wireless networks have become an indispensable part of our daily life, which have been widely used in civilian and military scenarios for communications. Security is a critical issue in wireless communications, since a large amount of important and private information is transferred over these wireless networks. Wireless communications are inherently public and visible in nature due to the open wireless medium, which is undesirable to preserve security and allows any unauthorized transceiver to detect, or eavesdrop on the wireless communications~\cite{bloch2011physical,Duong2016book}.
Against this background, conventional cryptography \cite{Menezes1996Handbook,Talbot2007Complexity} and information-theoretic physical layer security technologies \cite{yan2016artificial,hu2017artificial,youhong2017secure} have been developed to offer progressively higher levels of security by protecting the content of the message against eavesdropping. However, these technologies cannot mitigate the threat to a user's security and privacy from discovering the presence of the user or transmissions. Therefore, hiding a wireless transmission in the first place is widely demanded in some application scenarios. To meet this demand, covert communications have become a prominent solution to enable a wireless transmission between two users while guaranteeing a negligible probability of being detected by a warden \cite{bash2013limits,bash2015hiding,bloch2016covert,goeckel2016covert,wang2016fundamental}.

In the literature of covert communications, the authors of \cite{bash2013limits} demonstrated that $\mathcal{O}(\sqrt{n})$ bits of information can be transmitted to a legitimate receiver reliably and covertly in $n$ channel uses as $n \rightarrow \infty$ over additive white Gaussian noise (AWGN) channels, which is termed as square root law.
Following \cite{bash2013limits}, covert communications have been studied in different scenarios. For example, covert communications can be achieved when the warden has uncertainty about the receiver noise power~\cite{lee2015achieving,BiaoHe2017on}. In \cite{Sobers2015Covert,Sobers2017Covert}, the collaborative jammer comes to help to realize the covert communications. The effect of finite blocklength (i.e., a finite number of channel uses) over AWGN channels on covert communications was investigated in \cite{ShihaoYan2017Covert,Shihao2018Delay}. A covert communication system under block fading channels was examined in \cite{Shahzad2017Covert}, where transceivers have uncertainty on the related channel state information (CSI). In \cite{Hu2018ICC,Khurram2018fullduplex}, the authors utilized a full-duplex receiver to achieve covert communications in wireless fading channels and analyzed the covert communication limits. The covert communication with interference uncertainty from non-cooperative transmitters is studied in \cite{Xidian2018ICC}.

In some scenarios of wireless communication networks, a source node, instead of transmitting directly to a destination, transmits information to the destination with the aid of one of its neighbour nodes as a relay. Covert communications in the context of relay networks was examined in \cite{Hu2018covertrelay}, which showed that a relay can transmit confidential information to a corresponding destination covertly on top of forwarding the source's information to the destination.  Multi-hop covert communications over an arbitrary network in the presence of multiple collaborating wardens were investigated in \cite{Azadeh2018Multi}. With ubiquitous Internet of Things (IoT) devices (e.g., smart cities applications, intelligent transportation systems, wearable devices) adopted in everyday life, an unprecedented amount of connected objects and devices that store and exchange sensitive and confidential information such as real-time location and physiological information for e-health is transmitted over wireless channels. As such, crucial concerns on the security and privacy of wireless communications in IoT are emerging, which are believed to be the biggest barrier to the widespread adoption of IoT.
Covert communications can hide the existence of wireless transmissions and thus are able to address privacy issues in numerous applications of the emerging IoT.
In some practical application scenarios of IoT, a promising technique named wireless energy harvesting and information processing provides new opportunities and great convenience to solve the limited energy issues~\cite{Ali2013Relaying,He2015Harvest,YongZeng2015EH,YongZeng2017EH,Xiaoming2016Secrecy,qingqingwu2016EH,qingqingwu2016wirepower}.
With this technique, the source node can transfer energy through wireless communications to the relay and then requests the relay to help forwarding information to the destination. In this context, the energy and power are precious resources and thus the source node does not prefer or allow the relay node to use the harvested power for transmitting other information other than forwarding the source's information to the destination. As such, the relay's transmission of its own information should be kept covert from the source in order to guarantee the invisibility of this transmission.

In this work, the relay intends to transmit its own information to the destination covertly on top of forwarding the source's message, while source tries to detect this covert transmission to discover the illegitimate usage of the resource (the power obtained through the energy harvesting) allocated only for the purpose of forwarding the source's information. Specifically, we consider two existing energy harvesting strategies at the relay, namely the time switching (TS) and power splitting (PS) schemes, and aim to determine their performance in terms of the achievable effective covert rate from the relay to the destination subject to a specific covert communication constraint $\xi^{\ast} \geq 1 - \epsilon$, where $\xi^{\ast}$ denotes the minimum detection error probability at the source and $\epsilon$ is a small value specifying the required covertness. Note that in the TS scheme the signals used for energy harvesting and information processing at the relay are received from the source in different time slots, while in the PS scheme these signals are received simultaneously and then split into two streams with one stream processed by the energy receiver and the other processed by the information receiver. The main contributions of this work are summarized as follows.
\begin{itemize}
\item We first detail the strategies of transmitting covert information within the TS and PS schemes at the relay, focusing on determining the transmit power of the forwarded information and the covert information. Our analysis shows that in order to transmit its own information (i.e., covert information) without affecting forwarding the source's information to the destination, the relay has to harvest extra energy from the source. To this end, the relay has to adopt a more powerful energy harvester with a higher conversion efficiency factor when it transmits covert information relative to when it does not transmit covert information to the destination. The increase in the conversion efficiency factor represents a cost of the relay's covert transmission and thus we target at determining the amount of this increase in order to achieve the covert transmission limits from the relay to the destination for both the TS and PS schemes.

\item We develop the optimal detector at the source and derive its detection performance limit in terms of the minimum detection error probability for both the TS and PS schemes. Specifically, we first determine a sufficient test statistic at the source and construct a decision rule by comparing it to an arbitrary detection threshold. Then, we derive the false alarm and miss detection rates for any given detection threshold in closed-form expressions, based on which we analytically obtain the optimal detection threshold that minimizes the detection error probability. Our analysis indicates that the source's minimum detection error probabilities for the TS and PS schemes are exactly the same depending only on the ratio of the conversion efficiency factor $\eta_1$ of the energy harvester when the relay transmits covert information to the conversion efficiency factor $\eta_0$ when the relay does not transmit covert information.

\item We derive the required minimum conversion efficiency factor $\eta_1^{\ast}$ in a closed-form expression for any given $\eta_0$ in order to achieve the maximum effective covert rate $\Psi^{\ast}$ subject to $\xi^{\ast} \geq 1 - \epsilon$ for both the TS and PS schemes. Although the achieved maximum effective covert rates within these two schemes are different, our analysis demonstrates that the required $\eta_1^{\ast}$ is the same for these two schemes, which indicates that the cost of achieving $\Psi^{\ast}$ in terms of the conversion efficiency increase is the same for these two schemes. Our analysis also shows that the value of $\epsilon$ determines the limit on $\Psi^{\ast}$ and this limit depends on $\eta_0$ (i.e., the conversion efficiency factor when the relay does not transmit covert information) and $\eta_u$ (i.e., an upper bound on the conversion efficiency factor).
\end{itemize}

The rest of this paper is organized as follows. Section II details our system model and adopted assumptions. In Section III, we analyze the source's detection performance and the covert communication limits for the TS scheme. Section IV presents our analysis on the source's detection performance limits and covert communications with the PS scheme. Section V provides numerical results to thoroughly compare the TS and PS schemes, based on which we provide useful insights with regard to the impact of some system parameters on the achieved covert communications. Section VI draws conclusions.

\emph{Notation:} Scalar variables are denoted by italic symbols. Vectors are denoted by lower-case boldface symbols. Given a vector $\mathbf{x}$, $\mathbf{x}[i]$ denotes the $i$-th element of $\mathbf{x}$. Given a complex number, $|\cdot|$ denotes its modulus. $\mathbb{E}[\cdot]$ denotes expectation operation.

\section{System Model}
\subsection{Considered Scenario and Adopted Assumptions}

\begin{figure}[!t]
  \centering
  \includegraphics[scale=0.7]{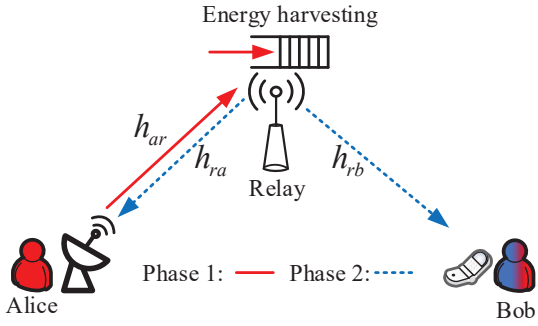}\\
  \caption{Covert communications with a wireless-powered relay.}\label{fig1}
\end{figure}

As shown in Fig.~\ref{fig1}, this work considers an one-way relay network with three nodes: the source (Alice), the relay, and the destination (Bob). Each node is equipped with a single antenna and operates in the half-duplex mode. We assume the wireless channels within our system model are subject to independent quasi-static Rayleigh fading with equal block length and the channel coefficients are independent and identically distributed (i.i.d.) circularly symmetric complex Gaussian random variables with zero-mean and unit-variance. We also assume that the direct link from Alice to Bob is not available due to blockage and the transmission from Alice to Bob has to be aided by the relay. Hence, the transmission from Alice to Bob occurs in two phases. In the first phase, Alice transmits energy and information to the relay. In the second phase, the self-sustained relay forwards the information from Alice to Bob with the harvested energy from Alice in the first phase. It is assumed that the total time of transmission is $T$.

The channel from Alice to the relay is denoted by $h_{ar}$ and the channel from the relay to Bob is denoted by $h_{rb}$. We assume that the relay knows both $h_{ar}$ and $h_{rb}$ perfectly. The energy consumption required for obtaining the CSI of $h_{ar}$ and $h_{rb}$ at the relay is assumed to be negligible compared to the power used for signal transmission \cite{Ali2013Relaying}. In practical wireless communication systems, CSI is usually obtained through CSI feedback from a receiver to a transmitter \cite{Xiaojuan2010On}. In this work, we assume the relay does not feed back CSI to Alice and thus seeks to transmit its own information to Bob on top of forwarding Alice's information, which should be kept covert from Alice, since Alice does not allow the relay to transmit its own information by using her energy. Therefore, in this work we assume that $h_{ra}$ is unknown to Alice.

\subsection{Transmission Strategy}

During the first phase of the cooperative transmission from Alice to Bob, the self-sustained relay harvests energy and receives the signals from Alice. We note that the procedures of energy harvesting and information processing are sequential and simultaneous for the TS and PS schemes, respectively.
In the second phase the relay forwards Alice's signals to Bob with the harvested energy, when the relay can decide whether to transmit its own message (covert message) to Bob on top of forwarding Alice's information. In this work, we assume that all the harvested energy at the relay will be used for transmission in the same block and the information (e.g., conversion efficiency) on the energy harvester that is used when the relay does not transmit covert information is publicly known. In this work, $\eta_0$ is the original energy efficiency when the relay does not desire covert transmission. However, when the relay intends to transmit covert message, the relay cannot conduct covert transmission if the energy efficiency is still $\eta_0$, since all the harvested energy is supposed to be used for forwarding the Alice's information to the destination. Hence, to conduct covert transmission, the relay has to use a more powerful energy harvester with a higher conversion efficiency $\eta_1$ such that it will have extra energy for potential covert transmission\cite{Hashemi2012Rectifier}. When the relay is to purchase the desired energy harvester, it has to determine what is the energy efficiency that can satisfy its covert transmission requirement. Intuitively, the higher the energy efficiency is, the easier the covert transmission can be constructed, since the relay can throw extra energy if it is not needed. However, a higher energy efficiency means a higher price for the energy harvester. Therefore, the relay wants to purchase a perfect energy harvester that can satisfy its covert transmission and also does not waste money on purchasing a too-good device. This work is to determine the energy efficiency of this perfect device. After purchasing this perfect device, the relay will always use this device and throw the extra harvested energy when it does not conduct covert transmission.

\subsection{Binary Detection at Alice}

Since Alice transmits energy to the relay in order to enable it forwarding Alice's information to Bob, the transmission of the relay will be monitored by Alice, who tries to detect the illegitimate usage of the harvested energy. To this end, Alice is to detect the wireless transmission of the covert information (the relay's own information to Bob, not the forwarded information) from the relay to Bob. Hence, Alice has a binary hypothesis detection problem, in which the relay does not transmit covert message to Bob in the null hypothesis $\Hnull$, while it does in the alternative hypothesis $\Halt$.
We define $P(\Hnull) = 1-\omega$ as the probability that Alice does not transmit and $P(\Halt) = \omega$ as the probability that Alice
transmits in time. $\mathcal{P}(\Hoalt|\Hnull)=\alpha$ is the false alarm rate and $\mathcal{P}(\Honull|\Halt)=\beta$ is the miss detection rate, while $\Hoalt$ and $\Honull$ are the binary decisions that infer whether the relay transmits covert message to Bob or not, respectively. The probability of error at Alice is given by
\begin{align} \label{xi}
\mathcal{P}_e&\triangleq  \mathcal{P}(\Hoalt|\Hnull)P(\Hnull) +  \mathcal{P}(\Honull|\Halt)P(\Halt)  \notag \\
&=(1-\omega)\alpha+\omega\beta.
\end{align}
The ultimate goal for Alice is to detect whether her observation comes from $\Hnull$ or $\Halt$ by applying a specific decision rule. The prior probabilities of hypotheses $\Hnull$ and $\Halt$ are assumed to be 1/2. Therefore, the covert communication constraint considered in this work is that the detection error probability at Alice should be no less than $1 - \epsilon$, i.e., $\xi \geq 1 - \epsilon$, where $\epsilon \in [0,1]$ is a predetermined value to specify the covert communication constraint\cite{bash2013limits,bash2016ignaore,Sobers2017Covert}, and the detection error probability $\xi$ is defined as
\begin{align} \label{xi}
\xi\triangleq  \alpha  +  \beta.
\end{align}

In the following two sections, we analyze the detection performance of Alice in the TS and PS schemes adopted by the relay to harvest energy from Alice, based on which the limits of covert transmissions from the relay to Bob are determined.

\section{Time Switching scheme}

In the TS scheme, $\phi$ is the fraction of the block time allocated to energy harvesting by the self-sustained relay from Alice, where we have $0<\phi< 1$. The remaining block time is divided into two equal parts, namely $(1-\phi)T/2$, for information transmissions from Alice to the relay and from the relay to Bob, respectively. In this section, we first detail these transmissions and then analyze the detection performance at Alice together with the performance of covert communications from the relay to Bob.

\subsection{Transmission from Alice to the Relay}
When Alice transmits signals, the received signal at relay is given by
\begin{align}\label{y_r_TS}
\mathbf{y}_r[i]=\sqrt{P_aL_{ar}}h_{ar}\mathbf{x}_a[i]+\mathbf{n}_{r,a}[i]+\mathbf{n}_{r,c}[i],
\end{align}
where $P_a$ is the transmit power of Alice, $L_{ar}\triangleq \nu (d_{ar})^{-m}$ is the path loss, $m$ is the path loss exponent, $\nu$ is a constant depending on carrier frequency, which is commonly set as $[c/(4\pi f_c)]^{2}$ with $c=3~\times10^8~\mathrm{m/s}$ and $f_c$ as the
carrier frequency \cite{Lifeng2017Wireless}, $d_{ar}$ is the distance from Alice to the relay, $\mathbf{x}_a$ is the signal transmitted by Alice satisfying $\mathbb{E}[\mathbf{x}_a[i]\mathbf{x}^{\dag}_a[i]]=1$, $i = 1, 2, \dots, n$ is the index of each channel use, $\mathbf{n}_{r,a}[i]$ is the antenna AWGN at the radio frequency (RF) band, with $\sigma^2_{r,a}$ as its variance, i.e., $\mathbf{n}_{r,a}[i] \thicksim\mathcal{CN}(0,\sigma^2_{r,a})$, and $\mathbf{n}_{r,c}[i]$ is the conversion AWGN due to the signal conversion from RF band to baseband with $\sigma^2_{r,c}$ as its variance, i.e., $\mathbf{n}_{r,c}[i] \thicksim\mathcal{CN}(0,\sigma^2_{r,c})$.
Note that the energy harvesting receiver rectifies the RF signal directly and gets the direct current to charge up the battery.
The fraction used for energy harvesting or information transmission is 1 during the block time. Therefore, the received noise at the relay is given by
\begin{align}
\sigma^2_r\triangleq \sigma^2_{r,a}+\sigma^2_{r,c}.
\end{align}
Then, the maximum energy the relay can possibly harvest (with the highest conversion efficiency factor 1) is given by
\begin{align}
E_{\mathrm{max}}^{\mathrm{EH}}=P_aL_{ar}|h_{ar}|^2 \phi T .
\end{align}
We note that this is not the actual amount of energy that can be harvested by the relay, which depends on the actual conversion efficiency factor.

\subsection{Transmission from the Relay to Bob}
In order to prevent Alice from canceling the component related to $\mathbf{x}_a$ at the procedure of detection, we assume that the relay and Bob share codebooks with Bob that are different from the codebooks used by Alice (Randomize-and-Forward strategy \cite{Dennis2011Artificial,Koyluoglu2012On,Chunxiao2013Secure}) or some specific secret keys such that the relay can modify its received signals from Alice (i.e., $\mathbf{x}_a$) before amplifying and forwarding them to Bob. For example, the relay can randomly delay the received signals in order to change their phases before the Amplify-and-Forward (AF) action. Following \eqref{y_r_TS}, we have the modified version of $\mathbf{y}_r$, which is given by
\begin{align}\label{y_r_TS_hat}
\hat{\mathbf{y}}_r[i]=\sqrt{P_aL_{ar}}h_{ar}\hat{\mathbf{x}}_a[i]+\hat{\mathbf{n}}_{r,a}[i]+\hat{\mathbf{n}}_{r,c}[i].
\end{align}

In some application scenarios of relay networks, utilizing the Randomize-and-Forward strategy~\cite{Dennis2011Artificial,Koyluoglu2012On,Chunxiao2013Secure} or some specific secret keys schemes, the source (Alice) and relay use different codebooks to transmit the message to the destination, such that an eavesdropper (if exists) cannot combine the received signals from the source and relay to decode the source's information. This in general will decrease the eavesdropper's eavesdropping ability and thus enhance the communication security in relay networks. Motivated by this, Alice is willing to allow this strategy adopted by the relay to protect each of individual links against potential eavesdroppers.

As the relay operates in the AF mode, it will forward a linearly amplified version of the received signal given in \eqref{y_r_TS_hat} to Bob. Therefore, the forwarded signal $\mathbf{x}_r[i]$ is given by
\begin{align}\label{fw_signal_TS}
\mathbf{x}_r[i]&=G\hat{\mathbf{y}}_r[i] \notag \\
&=G\left(\sqrt{P_aL_{ar}}h_{ar}\hat{\mathbf{x}}_a[i]+\hat{\mathbf{n}}_{r,a}[i]+\hat{\mathbf{n}}_{r,c}[i]\right),
\end{align}
where $G$ is a scaling scalar. In order to guarantee the power constraint at relay, the value of $G$ is chosen such that $\mathbb{E}[\mathbf{x}_r[i]\mathbf{x}^{\dag}_r[i]]=1$, which leads to
\begin{align} \label{G_TS}
G=\frac{1}{\sqrt{P_aL_{ar}|h_{ar}|^2+\sigma_r^2}}.
\end{align}

\subsubsection{Transmission of the Relay without Covert Information}

Under the null hypothesis $\Hnull$ (when the relay does not transmit its covert information to Bob), the relay only transmits $\mathbf{x}_r$ to Bob. Accordingly, the received signal at Bob is given by
\begin{align} \label{y_b_H0_TS}
\mathbf{y}_b[i]&=\sqrt{P_r^0L_{rb}}h_{rb}\mathbf{x}_r[i]+\mathbf{n}_{b,a}[i]+\mathbf{n}_{b,c}[i] \notag \\
&=\sqrt{P_r^0L_{rb}}h_{rb}G\Big(\sqrt{P_aL_{ar}}h_{ar}\hat{\mathbf{x}}_a[i]+\hat{\mathbf{n}}_{r,a}[i]+ \notag \\
&~~~~\hat{\mathbf{n}}_{r,c}[i]\Big)+\mathbf{n}_{b,a}[i]+\mathbf{n}_{b,c}[i],
\end{align}
where $L_{rb}\triangleq \nu (d_{rb})^{-m}$ is the path loss, $d_{rb}$ is the distance from relay to Bob, $\mathbf{n}_{b,a}[i]$ is the AWGN at Bob with $\sigma^2_{b,a}$ as its variance, i.e., $\sigma^2_{b,a}[i] \thicksim\mathcal{CN}(0,\sigma^2_{b,a})$, and $\mathbf{n}_{b,c}[i]$ is the sampled AWGN at Bob due to RF band to baseband signal conversion with $\sigma^2_{b,c}$ as its variance, i.e., $\sigma^2_{b,c}[i] \thicksim\mathcal{CN}(0,\sigma^2_{b,c})$. Hence, the total noise power at Bob is given by
\begin{align}\label{sigma_b2}
\sigma^2_b\triangleq\sigma^2_{b,a}+\sigma^2_{b,c}.
\end{align}
We note that in \eqref{y_b_H0_TS}, $P_r^0$ is the transmit power of $\mathbf{x}_r$ at the relay under $\Hnull$. Since all the harvested energy is used for transmission at the relay, $P_r^0$ is given by
\begin{align} \label{Pr0_TS}
P_r^0&=\frac{\eta_0 E_{\mathrm{max}}^{\mathrm{EH}}}{((1-\phi)T/2)} \notag \\
&=\frac{2\eta_0\phi P_a L_{ar}|h_{ar}|^2}{(1-\phi)},
\end{align}
where $\eta_0$ is the conversion efficiency factor of the energy harvester under $\Hnull$, which is publicly known.
Then, following \eqref{y_b_H0_TS} the SNR for $\mathbf{x}_a$ at Bob is derived as
\begin{align} \label{gamma0_TS}
\gamma_b^0&=\frac{P_r^0 L_{rb} |h_{rb}|^2 G^2 P_aL_{ar}|h_{ar}|^2}{P_r^0L_{rb}|h_{rb}|^2G^2\sigma_r^2+\sigma_b^2} .
\end{align}
The parameter $\phi$ that maximizes effective rate of $\mathbf{x}_a$ can be numerically obtained based on the method detailed in \cite{Ali2013Relaying} and thus it is assumed publicly known in this work. The value of $\phi$ is the same under $\Hnull$ and $\Halt$, since it should be agreed between Alice and the relay.

\subsubsection{Transmission of the Relay with Covert Information}

Under the alternative hypothesis $\Halt$ (when relay transmits the covert information to Bob on top of forwarding $\mathbf{x}_a$), the received signal at Bob is given by
\begin{align}\label{y_b_H1_TS}
\mathbf{y}_b[i]&=\sqrt{P_r^1L_{rb}}h_{rb}\mathbf{x}_r[i]+\sqrt{P_r^cL_{rb}}h_{rb}\mathbf{x}_c[i]+\mathbf{n}_b[i]  \notag \\
&=\sqrt{P_r^1L_{rb}}h_{rb}G\Big(\sqrt{P_aL_{ar}}h_{ar}\hat{\mathbf{x}}_a[i]+\hat{\mathbf{n}}_{r,a}[i]+ \notag \\
&~~~~\hat{\mathbf{n}}_{r,c}[i]\Big)+\sqrt{P_r^cL_{rb}}h_{rb}\mathbf{x}_c[i]+\mathbf{n}_{b,a}[i]+\mathbf{n}_{b,c}[i],
\end{align}
where $P_r^1$ is the relay's transmit power of $x_r$ under $\Halt$ and $P_r^c$ is the relay's transmit power of the covert information $\mathbf{x}_c$ satisfying $\mathbb{E}[\mathbf{x}_c[i]\mathbf{x}^{\dag}_c[i]]=1$. Again, since all the harvested energy is used for transmission in the relay, the total transmit power of relay is given by
\begin{align} \label{Pr1_plus_prc_TS}
P_r^1+P_r^c&=\frac{\eta_1 E_{\mathrm{max}}^{\mathrm{EH}}}{\left((1-\phi)T/2\right)} \notag \\
&=\frac{2\eta_1\phi P_a L_{ar} |h_{ar}|^2}{(1-\phi)},
\end{align}
where $\eta_1$ is the conversion efficiency factor of the energy harvester under $\Halt$. Considering the practical scenarios, we have $\eta_1\leq\eta_u<1$, where $\eta_u$ is the upper bound of conversion efficiency factor.
As previously mentioned, the covert transmission from relay to Bob should not affect the transmission from Alice to Bob. Therefore, here we assume that Bob always first decodes $\mathbf{x}_a$ with $\mathbf{x}_c$ as interference.
Following \eqref{y_b_H1_TS}, the signal-to-interference-plus-noise ratio (SINR) for $\mathbf{x}_a$ at Bob is derived as
\begin{align}\label{gamma1_TS}
\gamma_b^1&=\frac{P_r^1L_{rb}|h_{rb}|^2 G^2 P_aL_{ar}|h_{ar}|^2}{P_r^1L_{rb}|h_{rb}|^2 G^2\sigma_r^2+P_r^c L_{rb}|h_{rb}|^2+\sigma_b^2}.
\end{align}
We should guarantee that $\gamma_b^0=\gamma_b^1$ in order to avoid the impact of the covert transmission from the relay to Bob on the transmission from Alice to Bob. When $\gamma_b^0=\gamma_b^1$, we have
\begin{align}
&\frac{P_r^0 L_{rb} |h_{rb}|^2 G^2 P_aL_{ar}|h_{ar}|^2}{P_r^0L_{rb}|h_{rb}|^2G^2\sigma_r^2+\sigma_b^2} \notag \\
&=\frac{P_r^1L_{rb}|h_{rb}|^2 G^2 P_aL_{ar}|h_{ar}|^2}{P_r^1L_{rb}|h_{rb}|^2 G^2\sigma_r^2+P_r^c L_{rb}|h_{rb}|^2+\sigma_b^2} .
\end{align}
After some algebra manipulations, the above equation can be rewritten as
\begin{align} \label{Pr1_TS}
&P_r^1= \frac{P_r^0\left[\left(\frac{2\eta_1\phi P_a L_{ar} |h_{ar}|^2}{1-\phi}\right)L_{rb}|h_{rb}|^2+\sigma_b^2\right]}{P_r^0L_{rb}|h_{rb}|^2+\sigma_b^2} \\
&\!=\!\frac{2\eta_0\phi P_a L_{ar} |h_{ar}|^2\!\left[2\eta_1\phi P_a L_{ar}\!|h_{ar}|^2L_{rb} |h_{rb}|^2\!+\!(1\!-\!\phi)\sigma_b^2\right]\!}{(1\!-\!\phi)\left[2\eta_0\phi P_a L_{ar} |h_{ar}|^2 L_{rb} |h_{rb}|^2+(1\!-\!\phi)\sigma_b^2\right]}. \notag
\end{align}
Following \eqref{Pr0_TS} and \eqref{Pr1_TS}, we have
\begin{align}
P_r^1&=\frac{2\eta_1\phi P_aL_{ar}|h_{ar}|^2L_{rb}|h_{rb}|^2+(1-\phi)\sigma_b^2}{2\eta_0\phi P_a L_{ar}|h_{ar}|^2L_{rb}|h_{rb}|^2+(1-\phi)\sigma_b^2}P_r^0 \notag \\
&>P_r^0,
\end{align}
which confirms that the relay requires more power to forward $\mathbf{x}_a$ under $\Halt$ in order to guarantee the same end-to-end SNR/SINR of $\mathbf{x}_a$ at Bob under $\Hnull$. This is due to the fact that the covert information $\mathbf{x}_c$ causes interference at Bob for decoding $\mathbf{x}_a$. This indicates that the relay has to harvest more energy to support its covert transmission, which leads to $\eta_1 > \eta_0$. This means that the relay has to own an energy harvester with a higher conversion efficiency factor in order to conduct covert transmission on top of forwarding Alice's information to Bob. The increase in the conversion efficiency factor is a cost of the relay's covert transmission. In the following, we are going to determine the minimum value of $\eta_1$ to achieve the covert communication limits from the relay to Bob.
Intuitively, the relay can purchase an energy harvester with the highest conversion efficiency factor 1 and only use partial of the harvested energy to perform covert transmission in order to guarantee the covert communication constraint (i.e., $\xi^{\ast} \geq 1 - \epsilon$). We note that a higher  conversion efficiency factor means a higher cost and thus in this work we focus on the minimum value of $\eta_1$ that achieves the covert communication limits that indicates the lowest cost of the relay's covert transmission. Following \eqref{Pr1_plus_prc_TS} and \eqref{Pr1_TS}, the transmit power of $\mathbf{x}_c$ at the relay is given by
\begin{align} \label{Prc_TS}
P_r^c&=\frac{2\eta_1\phi P_a L_{ar}|h_{ar}|^2}{1-\phi}-P_r^1  \notag \\
&=\frac{2(\eta_1-\eta_0)\phi P_aL_{ar}|h_{ar}|^2\sigma_b^2}{2\eta_0\phi P_aL_{ar}|h_{ar}|^2L_{rb}|h_{rb}|^2+(1-\phi)\sigma_b^2}.
\end{align}

\subsection{Detection Performance and Optimal Detection Threshold at Alice}

In this subsection, we present the optimal detection strategy at Alice and her detection performance limits. To this end, we first determine a sufficient test statistic at Alice, based on which we construct a decision rule for an arbitrary detection threshold. Then, we derive the detection performance in terms of the false alarm and miss detection rates for any given detection threshold. Finally, we analytically obtain the optimal detection threshold that minimizes the detection error probability.

When the relay transmits to Bob, Alice will detect whether the relay transmits $\mathbf{x}_c$ on top of forwarding $\mathbf{x}_a$ to Bob. We recall that the relay does not transmit $\mathbf{x}_c$ in the null hypothesis $\Hnull$ while it does in the alternative hypothesis $\Halt$. Then, the received signal at Alice (when the relay transmits signals) is given by
\begin{align}
\mathbf{y}_a[i]\!=\!\left\{\!
  \begin{array}{ll}
    \sqrt{\!P_r^0L_{ra}}h_{ra}\mathbf{x}_r[i]\!+\!\mathbf{n}_{a}[i], &\Hnull, \\
    \sqrt{\!P_r^1L_{ra}}h_{ra}\mathbf{x}_r[i]\!+\!\sqrt{\!P_r^cL_{ra}}h_{ra}\mathbf{x}_c[i]\!+\!\mathbf{n}_{a}[i],  &\Halt,
  \end{array}
\right. \label{ya}
\end{align}
where $L_{ra}$ is the path loss from the relay to Alice, $\mathbf{n}_{a}[i]$ is the AWGN at Alice with $\sigma^2_{a}$ as its variance, i.e., $\sigma^2_{a}[i] \thicksim\mathcal{CN}(0,\sigma^2_{a})$.

\begin{lemma}\label{lemma1}
Under a specified condition that the relay has modified the forwarded message (by using randomize-and-forward strategy or some secrecy keys), Alice employs a radiometer as the detection test and the radiometer is demonstrated to be the optimal detector.
\end{lemma}

\begin{IEEEproof}
This proof is provided in Appendix \ref{app:proof optimal detector}
\end{IEEEproof}

In this work we assume that, which the. In the TS scheme, as per Appendix \ref{app:proof optimal detector}, the sufficient statistic $T$ is given by
\begin{align}\label{t_a_TS}
T &\overset{a}{=}\left\{\!
  \begin{array}{ll}
    \frac{2\eta_0\phi P_a L_{ar}^2|h_{ar}|^4}{1-\phi} + \sigma_a^2, &~~~\Hnull, \\
    \frac{2\eta_1\phi P_a L_{ar}^2|h_{ar}|^4}{1-\phi} + \sigma_a^2,  &~~~\Halt,
  \end{array}
\right.
\end{align}
and $\overset{a}{=}$ is achieved by using \eqref{Pr0_TS} and \eqref{Pr1_plus_prc_TS}.

Then, we derive the false alarm and miss detection rates at Alice for an arbitrary threshold $\tau$ in the following theorem, based on which we will tackle the optimization of $\tau$ in Theorem~\ref{theorem2}.
\begin{theorem}\label{theorem1}
The false alarm and miss detection rates at Alice for an arbitrary detection threshold $\tau$ are, respectively, derived as
\begin{align}
\alpha  &=\left\{
  \begin{array}{ll}
    1,  &~~~~~\tau<\sigma_a^2, \\
    \exp\left\{-\frac{1}{\lambda_{ar}}\sqrt{\frac{(\tau-\sigma_a^2)(1-\phi)}{2\eta_0\phi P_a L_{ar}^2}}\right\},  &~~~~~\tau>\sigma_a^2,
  \end{array} \label{PFA_TS}
\right.  \\
\beta&=\left\{
  \begin{array}{ll}
    0,  &\tau<\sigma_a^2, \\
    1-\exp\left\{-\frac{1}{\lambda_{ar}}\sqrt{\frac{(\tau-\sigma_a^2)(1-\phi)}{2\eta_1\phi P_aL_{ar}^2}}\right\},  &\tau>\sigma_a^2.
  \end{array} \label{PMD_TS}
\right.
\end{align}
\end{theorem}

\begin{IEEEproof}
For a given $\tau$, following~\eqref{t_a_TS}, the false alarm rate and miss detection rate are, respectively, given by
\begin{align}
\alpha &=\mathcal{P}\left[\frac{2\eta_0\phi P_a L_{ar}^2 |h_{ar}|^4}{1-\phi} + \sigma_a^2\geq\tau\right]  \notag \\
&=\left\{
  \begin{array}{ll}
    1,  &~~~~\tau<\sigma_a^2, \\
    \mathcal{P}\left[|h_{ar}|^4 \geq \frac{(\tau-\sigma_a^2)(1-\phi)}{2\eta_0\phi P_a L_{ar}^2}\right],  &~~~~\tau>\sigma_a^2,
  \end{array} \label{PFA_2_TS}
\right.\\
\beta &=\mathcal{P}\left[\frac{2\eta_1\phi P_a L_{ar}^2 |h_{ar}|^4}{1-\phi} + \sigma_a^2<\tau\right] \notag \\
&=\left\{
  \begin{array}{ll}
    0,  &~~~~\tau<\sigma_a^2, \\
    \mathcal{P}\left[|h_{ar}|^4 < \frac{(\tau-\sigma_a^2)(1-\phi)}{2\eta_1\phi P_a L_{ar}^2}\right],  &~~~~\tau>\sigma_a^2.
  \end{array} \label{PMD_2_TS}
\right.
\end{align}
Considering quasi-static Rayleigh fading, the cumulative distribution function (CDF) of $|h_{ar}|^4$ is given by $F_{|h_{ar}|^4}(x) = 1 - \exp(-\sqrt{x}/\lambda_{ar})$. Following \eqref{PFA_2_TS} and \eqref{PMD_2_TS}, we achieve the desired results in \eqref{PFA_TS} and \eqref{PMD_TS} after some algebra manipulations.
\end{IEEEproof}

\begin{theorem}\label{theorem2}
The optimal threshold that minimizes $\xi$ is derived as
\begin{align} \label{tau_ast_TS}
\tau^{\ast}=\sigma_a^2+\frac{1}{1-\phi}\left[\frac{\lambda_{ar}\sqrt{2\phi P_aL_{ar}^2\eta_0\eta_1}}{2(\sqrt{\eta_1}-\sqrt{\eta_0})}\ln\left(\frac{\eta_1}{\eta_0}\right)\right]^2.
\end{align}
\end{theorem}

\begin{IEEEproof}
Following \eqref{PFA_TS} and \eqref{PMD_TS}, we have the detection error probability at Alice as
\begin{align} \label{xi_TS}
\xi=\left\{
  \begin{array}{ll}
    1, & \tau \leq \sigma_a^2, \\
    1+\exp\left\{-\frac{1}{\lambda_{ar}}\sqrt{\frac{(\tau-\sigma_a^2)(1-\phi)}{2\eta_0\phi P_a L_{ar}^2}}\right\}-\\
    \exp\left\{-\frac{1}{\lambda_{ar}}\sqrt{\frac{(\tau-\sigma_a^2)(1-\phi)}{2\eta_1\phi P_a L_{ar}^2}}\right\},  &\tau>\sigma_a^2.
  \end{array}
\right.
\end{align}
As per \eqref{xi_TS}, Alice will not set $\tau \leq \sigma_a^2$, since $\xi = 1$ is the worst case for Alice. Following \eqref{xi_TS}, we derive the first derivative of $\xi$ with respect to $\tau$ for $\tau>\sigma_a^2$ as
${\partial\xi}/{\partial\tau}=\kappa_1(\tau)\kappa_2(\tau)$,
where
\begin{align}
\kappa_1(\tau)&\!\triangleq\! \frac{\sqrt{1-\phi}\exp\left(-\frac{\sqrt{(\tau-\sigma_a^2)(1-\phi)}}{\lambda_{ar}\sqrt{2\eta_1\phi P_a L_{ar}^2}}\right)}{2\lambda_{ar}\sqrt{2\eta_1\phi P_aL_{ar}^2(\tau-\sigma_a^2)}},  \label{kappa3_tau}  \\
\kappa_2(\tau)&\!\triangleq\! 1\!-\!\sqrt{\frac{\eta_1}{\eta_0}}\exp\!\left[\!-\frac{\sqrt{(\tau\!-\!\sigma_a^2)(1\!-\!\phi)}}{\lambda_{ar}\sqrt{2\phi P_a L_{ar}^2}}\left(\frac{1}{\sqrt{\eta_0}}\!-\!\frac{1}{\sqrt{\eta_1}}\!\right)\!\right].  \label{kappa4_tau}
\end{align}
We note that $\kappa_1(\tau)>0$ due to $\tau>\sigma_a^2$. Hence, the value of $\tau$ that ensures ${\partial \xi}/{\partial\tau}=0$ is the one guarantees $\kappa_2(\tau) = 0$, which is given by
\begin{align} \label{tau_dag_definition_TS}
\tau^{\dag}=\sigma_a^2+\frac{1}{1-\phi}\left[\frac{\lambda_{ar}\sqrt{2\phi P_aL_{ar}^2\eta_0\eta_1}}{2(\sqrt{\eta_1}-\sqrt{\eta_0})}\ln\left(\frac{\eta_1}{\eta_0}\right)\right]^2.
\end{align}
We note that ${\partial \xi}/{\partial\tau}<0$, for $\tau<\tau^{\dag}$, and ${\partial \xi}/{\partial\tau}>0$, for $\tau>\tau^{\dag}$. This is due to the fact that the term $\kappa_2(\tau)$ given in \eqref{kappa4_tau} is monotonically increasing with respect to $\tau$. Noting that $\tau^{\dag}>\sigma_a^2$, we have that $\tau^{\dag}$ minimizes $\xi$ for $\tau>\sigma_a^2$. Noting $\xi$ is a continuous function of $\tau$, we can conclude that the optimal threshold is $\tau^{\dag}$.
\end{IEEEproof}

\begin{corollary}\label{corollary1}
The minimum value of the detection error probability $\xi$ at Alice is
\begin{align} \label{xi_ast_TS}
\xi^{\ast}=1-\varphi^{\frac{1}{2(1-\sqrt{\varphi})}}\left(\frac{1}{\sqrt{\varphi}}-1\right),
\end{align}
where the system overhead $\varphi$ is defined as the ratio of the power without covert transmission ($P_r^0$) to the total power with covert transmission ($P_r^1+P_r^c$).
\begin{align}
\varphi=\frac{P_r^0}{P_r^1+P_r^c}=\frac{\eta_0}{\eta_1}.
\end{align}
\end{corollary}

\begin{IEEEproof}
Substituting $\tau^{\ast}$ into \eqref{xi_TS}, we obtain the minimum value of $\xi$, which is given by
\begin{align} \label{xi_ast_raw_TS}
\xi^{\ast}&=1+\exp\left\{-\frac{\sqrt{\eta_1}}{2(\sqrt{\eta_1}-\sqrt{\eta_0})}\ln\left(\frac{\eta_1}{\eta_0}\right)\right\} - \notag \\
&~~~~\exp\left\{-\frac{\sqrt{\eta_0}}{2(\sqrt{\eta_1}-\sqrt{\eta_0})}\ln\left(\frac{\eta_1}{\eta_0}\right)\right\}.
\end{align}
Following \eqref{xi_ast_raw_TS}, we achieve the desired result in \eqref{xi_ast_TS} after some algebra manipulations.
\end{IEEEproof}

\begin{remark} \label{remark1}
Corollary~\ref{corollary1} indicates that the minimum detection error probability $\xi^{\ast}$ only depends on $\varphi = \eta_0/\eta_1$ rather than any other system parameters. This is a surprising result and is due to the fact that the impact of other system parameters has been eliminated by the optimal detection threshold, which is confirmed by that the false alarm and miss detection rates derived in Theorem~1 together with the optimal detection threshold achieved in Theorem~2 are functions of the other system parameters.
\end{remark}

\begin{remark} \label{remark2}
The minimum detection error probability $\xi^{\ast}$ is a  monotonically increasing function of $\varphi$, which is confirmed by
\begin{align} \label{partial_xi_ast}
\frac{\partial\xi^{\ast}}{\partial\varphi}=-\frac{\ln(\varphi)}{4\varphi(1-\sqrt{\varphi})}\varphi^{\frac{1}{2(1-\sqrt{\varphi})}}>0.
\end{align}
This means that for a fixed $\eta_0$, $\xi^{\ast}$ decreases as $\eta_1$ increases (it becomes easier for Alice to detection the relay's covert transmission). Intuitively, this is due to that as $\eta_1$ increases the relay will transmit the covert information with a higher power, since all the harvested energy is used for transmission at the relay.
\end{remark}

\begin{corollary}\label{corollary2}
The value range of $\xi^{\ast}$ is $\left[1-(\eta_0/\eta_u)^{{\sqrt{\eta_u}}/(2(\sqrt{\eta_u}-\sqrt{\eta_0}))}(\sqrt{\eta_u}/\sqrt{\eta_0}-1),1\right]$, where $\eta_u$ is the upper bound of conversion efficiency factor.
\end{corollary}
\begin{IEEEproof}
Noting $\varphi=\eta_0/\eta_1$,  for a given $\eta_0$ the minimum value of $\varphi$ is achieved when $\eta_1=\eta_u$. Following Remark~\ref{remark2}, the minimum value of $\xi^{\ast}$ in \eqref{xi_ast_TS} is given by
\begin{align}
\xi^{\ast}\left(\varphi=\frac{\eta_0}{\eta_u}\right)=1-\left(\frac{\eta_0}{\eta_u}\right)^{\frac{\sqrt{\eta_u}}{2(\sqrt{\eta_u}-\sqrt{\eta_0})}}\left(\frac{\sqrt{\eta_u}}{\sqrt{\eta_0}}-1\right).
\end{align}
The maximum value of $\varphi$ is achieved when $\eta_1=\eta_0$. Then, using L'Hospital's rule, we can obtain the maximum value of $\xi^{\ast}$ as $\varphi \rightarrow 1$, as shown below:
\begin{align} \label{xi_ast__lim_1}
\lim_{\varphi \rightarrow 1} {\xi^{\ast}(\varphi)}&=1+\lim_{\varphi \rightarrow 1} \exp\left\{\frac{\ln\left(\varphi\right)}{2(1-\sqrt{\varphi})}\right\}- \notag \\
&~~~~\lim_{\varphi \rightarrow 1}  \exp\left\{\frac{\sqrt{\varphi}\ln\left(\varphi\right)}{2(1-\sqrt{\varphi})}\right\} \notag \\
& =1.
\end{align}
This completes the proof of Corollary~\ref{corollary2}.
\end{IEEEproof}

\subsection{Optimization of Covert Transmission}

When the relay transmits covert information, Bob first decodes $\mathbf{x}_a$ and then subtracts the corresponding component from its received signal $\mathbf{y}_b$ given in \eqref{y_b_H1_TS} in order to decode the covert information $\mathbf{x}_c$.
Therefore, the effective received signal used to decode $\mathbf{x}_c$ is given by
\begin{align} \label{tilde y_b_H1}
\tilde{\mathbf{y}}_b[i]&=\sqrt{P_r^cL_{rb}}h_{rb}\mathbf{x}_c[i]+\sqrt{P_r^1L_{rb}}h_{rb}G\Big(\hat{\mathbf{n}}_{r,a}[i]+ \notag \\
&~~~~\hat{\mathbf{n}}_{r,c}[i]\Big)+\mathbf{n}_{b,a}[i]+\mathbf{n}_{b,c}[i].
\end{align}
Then, following \eqref{tilde y_b_H1} the SNR for $\mathbf{x}_c$ at Bob is
\begin{align} \label{gamma_c_TS}
\gamma_c &= \frac{P_r^cL_{rb}|h_{rb}|^2}{P_r^1L_{rb}|h_{rb}|^2 G^2\sigma^2_r +\sigma^2_b} \notag \\
&\overset{b}{=} \frac{(Q_2 - Q_1)\sigma_b^2}{\frac{Q_1[Q_2+ (1 - \phi)\sigma_b^2)]\sigma_r^2}{(1 - \phi)(P_a L_{ar}|h_{ar}|^2 + \sigma_r^2)} + \left[Q_1 + (1 - \phi)\sigma_b^2\right]\sigma_b^2},\notag
\end{align}
where
\begin{align}
Q_1&\triangleq 2 \eta_0\phi P_a L_{ar}|h_{ar}|^2 L_{rb}|h_{rb}|^2,   \notag \\
Q_2&\triangleq 2 \eta_1\phi P_a L_{ar}|h_{ar}|^2 L_{rb}|h_{rb}|^2,
\end{align}
and $\overset{b}{=}$ is obtained based on \eqref{G_TS}, \eqref{Pr1_TS}, and \eqref{Prc_TS}.
Following \eqref{gamma_c_TS} and considering Rayleigh fading for $h_{ar}$ and $h_{rb}$, the average rate of the covert transmission from the relay to Bob is given by
\begin{align} \label{rate_TS}
&C\!=\!\int_0^{\infty}\int_0^{\infty}\log_2\left\{1+\gamma_c\right\}f_{|h_{ar}|^2}(x)f_{|h_{rb}|^2}(y)\mathrm{d}x \mathrm{d}y \notag \\
&\!=\!\frac{1}{\lambda_{ar}\lambda_{rb}}\int_0^{\infty}\int_0^{\infty}\exp{\left[-\left(\frac{x}{\lambda_{ar}}+\frac{y}{\lambda_{rb}}\right)\right]}\log\Bigg\{1+ \notag \\
&\frac{[Q_2(x,y)-Q_1(x,y)]\sigma_b^2}{\frac{Q_1\!(x,y)[Q_2\!(x,y)+(1-\phi)\sigma_b^2]\sigma_r^2}{(1-\rho)P_aL_{ar}x+\sigma_r^2}\!+\! [Q_1\!(x,y)\!+\!(1\!-\!\phi)\sigma_b^2]\sigma_b^2 }\!\Bigg\}\mathrm{d}x\mathrm{d}y.
\end{align}
Since $(1-\phi)T/2$ is the effective communication time between the relay and Bob in the total block time $T$, for the TS scheme the effective covert rate is defined as
\begin{align} \label{thoughput_TS}
\Psi&=\frac{[(1-\phi)T/2]}{T}C  \notag \\
&=\frac{(1-\phi)}{2}C.
\end{align}

In this work, we consider $\eta_1$ as the only system parameter of interest, which represents the conversion efficiency factor under $\Halt$ (where the relay transmits covert information on top of forwarding Alice's messages to Bob) and thus indicates the cost of the relay's covert transmission. Hence, the optimization problem at relay of maximizing the effective covert rate subject to a certain covert communication constraint is given by
\begin{equation}\label{P1_TS}
\begin{aligned}
\underset{\eta_0 \leq \eta_1 \leq \eta_u}{\max} \quad &\Psi \\
\text{s. t.} \quad  &\xi^{\ast}(\varphi) \geq 1 -\epsilon.
\end{aligned}
\end{equation}
The maximum value of $\Psi$ is then achieved by substituting the optimal value of $\eta_1$ (which is derived in the following theorem) into \eqref{thoughput_TS}, which is denoted by $\Psi^{\ast}$.

\begin{theorem} \label{theorem3}
For a given conversion efficiency factor $\eta_0$ under $\Hnull$, the optimal value (i.e., minimum value) of $\eta_1$ that maximizes the effective covert rate $\Psi$ subject to the covert communication constraint $\xi^{\ast}(\varphi) \geq 1 - \epsilon$ is given by
\begin{align} \label{eta1_ast_TS}
\eta_1^{\ast}=\left\{
  \begin{array}{ll}
    \frac{\eta_0}{\varphi_{\epsilon}},  &\epsilon\leq \left(\frac{\eta_0}{\eta_u}\right)^{\frac{\sqrt{\eta_u}}{2(\sqrt{\eta_u}-\sqrt{\eta_0})}}\left(\sqrt{\frac{\eta_u}{\eta_0}}-1\right), \\
    \eta_u,  &\mathrm{otherwise},
  \end{array}
\right.
\end{align}
where $\varphi_{\epsilon}$ is the solution of $\varphi$ to $\xi^{\ast}(\varphi)=1-\epsilon$ and $\xi^{\ast}(\varphi)$ can be obtained as per \eqref{xi_ast_TS}.
\end{theorem}

\begin{IEEEproof}
As mentioned in Remark~\ref{remark2}, the minimum detection error probability $\xi^{\ast}$ is a monotonically increasing function of $\varphi$. Therefore, we have $\eta_1\leq \varphi_{\epsilon}/\eta_0$ in order to guarantee $\xi^{\ast}(\varphi) \geq 1 - \epsilon$. Following \eqref{gamma_c_TS}, $\gamma_c$ can be rewritten as a function of $\eta_1$, which is given by
\begin{align} \label{gamma_c_TS2}
\gamma_c = \frac{B_1(1-\eta_0/\eta_1)}{B_2+B_3/\eta_1},
\end{align}
where
\begin{align} \label{B1B2B3}
B_1&\!\triangleq\! 2(1\!-\!\phi)\phi P_a L_{ar}|h_{ar}|^2L_{rb}|h_{rb}|^2\sigma_b^2(P_aL_{ar}|h_{ar}|^2\!+\!\sigma_r^2) , \notag \\
B_2&\!\triangleq\! 4\eta_0\phi^2 P_a^2L_{ar}^2|h_{ar}|^4 L_{rb}|h_{rb}|^2,\notag \\
B_3&\!\triangleq\! (1-\phi)\sigma^2_b\Big[2\eta_0\phi P_aL_{ar}|h_{ar}|^2L_{rb}|h_{rb}|^2\sigma^2_r+  \notag \\
&~~~~P_aL_{ar}|h_{ar}|^2+\sigma_r^2\Big].
\end{align}
Following \eqref{gamma_c_TS2} and noting that the terms (i.e., $B_1$, $B_2$, and $B_3$) in \eqref{B1B2B3} are no less than 0, we can conclude that $\gamma_c$ is monotonically increasing with respect to $\eta_1$ for any value of $|h_{ar}|^2$ and $|h_{rb}|^2$. Then, following the Leibniz integral rule we know that $\Psi$ is a monotonically increasing function of $\eta_1$. Hence, the optimal value of $\eta_1$ to the optimization problem given in \eqref{P1_TS} without the constraint $\eta_1 \leq \eta_u$ is the one that guarantees $\xi^{\ast}(\varphi) = 1 - \epsilon$.
Considering $\eta_1 \leq \eta_u$, we finally have the desired result as given in \eqref{eta1_ast_TS}.
\end{IEEEproof}

\begin{remark} \label{remark3}
Following Theorem~\ref{theorem3}, we note that when $\eta_1^{\ast} = \eta_0/\varphi_{\epsilon}$, it is the covert communication constraint that limits the effective covert rate, since $\varphi_\epsilon$ is determined only by $\epsilon$ rather than any other system parameters. When $\eta_1 = \eta_u$, it is the energy harvester that limits the effective covert rate, since the relay cannot harvest more energy to conduct the covert transmission (although it is allowed to do that, i.e., $\xi^{\ast}(\varphi) \geq 1 - \epsilon$ can still be guaranteed if the relay transmits covert information with higher power).
\end{remark}

\begin{remark} \label{remark4}
If the relay sets $\eta_1=\eta_u$ when $\eta_1^{\ast} = \eta_0/\varphi_{\epsilon}<\eta_u$ and only uses partial of the harvested energy to perform covert transmission, it can still guarantee the covert communication constraint $\xi^{\ast}(\varphi) \geq 1 -\epsilon$. However, this means that the relay uses an energy harvester with the highest conversion efficiency factor 1 rather than the required minimum conversion efficiency factor $\eta_1^{\ast}$, which also means that relay wastes some harvested energy and wastes some cost on operating a better energy harvester than the necessary one.
\end{remark}

\section{Power Splitting Scheme}

In the PS scheme, half of the block time (i.e., $T/2$) is used for simultaneous information and power transfer from Alice to the relay and the remaining $T/2$ block time is used for information transmission from the relay to Bob. In order to harvest energy, the relay splits the received signals from Alice into two fractions during the first $T/2$ block time. In this work, we denote the fraction used for energy harvesting as $\rho$ and the remaining fraction $1 - \rho$ is used for information transmission, where we have $0 < \rho < 1$.

\subsection{Transmission from Alice to the Relay}

After the signal splitting, the received signal at the relay for information delivery is given by
\begin{align}\label{y_r_PS}
\mathbf{y}_r[i]\!=\!\sqrt{\!P_aL_{ar}(1\!-\!\rho)}h_{ar}\mathbf{x}_a[i]\!+\!\sqrt{\!(1\!-\!\rho)}\mathbf{n}_{r,a}[i]\!+\!\mathbf{n}_{r,c}[i].
\end{align}
and the fraction used for energy harvesting is denoted by $\rho$ and the remaining fraction $1 - \rho$ is used for information transmission. Note that the energy harvesting happens at RF band and only $1 - \rho$ of the total power is used for information transmission, and therefore the total noise power is given by
\begin{align} \label{sigma_r2_PS}
\sigma^2_r\triangleq (1-\rho)\sigma^2_{r,a}+\sigma^2_{r,c}.
\end{align}
Then, the maximum energy the relay can possibly harvest (with the highest conversion efficiency factor 1) is given by
\begin{align}\label{y_r_EH}
E_{\mathrm{max}}^{\mathrm{EH}}=\rho P_aL_{ar}|h_{ar}|^2 (T/2).
\end{align}
Again, it should be noted that this is not the actual amount of energy that can be harvested by the relay, which depends on the actual conversion efficiency factor.

\subsection{Transmission from the Relay to Bob}

In this subsection, we detail the transmission strategies of the relay when it does and does not transmit covert information to Bob in the PS scheme. Similar to \eqref{y_r_TS_hat}, the $\mathbf{y}_r$ is modified to $\hat{\mathbf{y}}_r$.  As a result, the forwarded signal $\mathbf{x}_r[i]$ at the relay is given by
\begin{align}\label{x_r}
\mathbf{x}_r[i]&=G\hat{\mathbf{y}}_r[i] \notag \\
&=G\Big(\sqrt{P_aL_{ar}(1-\rho)}h_{ar}\hat{\mathbf{x}}_a[i]+ \notag \\
&~~~~\sqrt{(1-\rho)}\hat{\mathbf{n}}_{r,a}[i]
+\hat{\mathbf{n}}_{r,c}[i]\Big).
\end{align}
In order to guarantee the power constraint at relay, the value of $G$ is chosen such that $\mathbb{E}[\mathbf{x}_r[i]\mathbf{x}^{\dag}_r[i]]=1$, which leads to
\begin{align} \label{G_PS}
G=\frac{1}{\sqrt{(1-\rho)P_aL_{ar}|h_{ar}|^2+\sigma^2_r}}.
\end{align}

\subsubsection{Transmission of the Relay without Covert Information}

Under $\Hnull$, the relay does not transmit covert information and only transmits $\mathbf{x}_r$ to Bob. Accordingly, the received signal at Bob is given by
\begin{align}\label{y_b_H0}
\mathbf{y}_b[i]&=\sqrt{P_r^0L_{rb}}h_{rb}\mathbf{x}_r[i]+\mathbf{n}_{b,a}[i]+\mathbf{n}_{b,c}[i] \notag \\
&=\sqrt{P_r^0L_{rb}}h_{rb}G_0\Big(\sqrt{P_aL_{ar}(1-\rho)}h_{ar}\hat{\mathbf{x}}_a[i]+ \notag \\
&~~~~~\sqrt{(1-\rho)}\hat{\mathbf{n}}_{r,a}[i]+\hat{\mathbf{n}}_{r,c}[i]\Big)+\mathbf{n}_{b,a}[i]+\mathbf{n}_{b,c}[i],
 \end{align}
where we recall that $\mathbf{n}_{b,a}[i]$ and $\mathbf{n}_{b,c}[i]$ are defined in \eqref{sigma_b2} and $P_r^0$ is the transmit power of $\mathbf{x}_r$ under $\Hnull$, which is given by
\begin{align} \label{Pr0_PS}
P_r^0&=\frac{\eta_0 E_{\mathrm{max}}^{\mathrm{EH}}}{(T/2)} \notag \\
&=\eta_0\rho P_aL_{ar} |h_{ar}|^2.
\end{align}
Then, following \eqref{y_b_H0}, the SNR for $\mathbf{x}_a$ under $\Hnull$ at Bob is derived as
\begin{align} \label{gamma0_PS}
\gamma_b^0&=\frac{P_r^0L_{rb}|h_{rb}|^2 G^2 P_a(1-\rho)L_{ar}|h_{ar}|^2}{P_r^0L_{rb}|h_{rb}|^2 G^2 \sigma^2_r+\sigma_b^2}  .
\end{align}
The parameter $\rho$ that maximizes effective rate of $\mathbf{x}_a$ can be numerically obtained based on the method detailed in \cite{Ali2013Relaying} and thus it is assumed publicly known in this work. We note that $\rho$ can be different under $\Hnull$ and $\Halt$ in the PS scheme, which is different from the TS scheme, since the value of $\rho$ is solely determined by the relay. However, in this work we do not consider different values of $\rho$ under $\Hnull$ and $\Halt$ in order to seek a fair comparison between the TS and PS schemes.

\subsubsection{Transmission of the Relay with Covert Information}

Under $\Halt$, the relay transmits covert information to Bob on top of forwarding $\mathbf{x}_a$ and thus the received signal at Bob is given by
\begin{align}\label{y_b_H1}
\mathbf{y}_b[i]&\!=\!\sqrt{\!P_r^1L_{rb}}h_{rb}\mathbf{x}_r[i]\!+\!\sqrt{\!P_r^cL_{rb}}h_{rb}\mathbf{x}_c[i]\!+\!\mathbf{n}_{b,a}[i]\!+\!\mathbf{n}_{b,c}[i] \notag \\
&=\sqrt{P_r^1L_{rb}}h_{rb}G\Big(\sqrt{P_aL_{ar}(1-\rho)}h_{ar}\hat{\mathbf{x}}_a[i]+ \notag \\
&~~~\sqrt{(1-\rho)}\hat{\mathbf{n}}_{r,a}[i]+\hat{\mathbf{n}}_{r,c}[i]\Big)+ \sqrt{P_r^cL_{rb}}h_{rb}\mathbf{x}_c[i]+ \notag \\
&~~~\mathbf{n}_{b,a}[i]+\mathbf{n}_{b,c}[i],
\end{align}
where $P_r^1$ is the relay's transmit power of $x_r$ in this case and $P_r^c$ is the relay's transmit power of $\mathbf{x}_c$ satisfying $\mathbb{E}[\mathbf{x}_c[i]\mathbf{x}^{\dag}_c[i]]=1$. As assumed, all the harvested energy at the relay is used for transmission and thus the total transmit power of the relay is given by
\begin{align} \label{Pr1_plus_prc_PS}
P_r^1+P_r^c&=\frac{\eta_1 E_{\mathrm{max}}^{\mathrm{EH}}}{T/2} \notag \\
&=\eta_1\rho P_a L_{ar}|h_{ar}|^2,
\end{align}
where $\eta_1$ is the conversion efficiency factor of the energy harvester adopted under $\Halt$.
Following \eqref{y_b_H1}, the SINR for $\mathbf{x}_a$ at Bob is derived as
\begin{align}\label{gamma1_PS}
\gamma_b^1=\frac{P_r^1L_{rb}|h_{rb}|^2G^2P_a(1-\rho)L_{ar}|h_{ar}|^2}{P_r^1L_{rb}|h_{rb}|^2G^2\sigma^2_r+P_r^cL_{rb}|h_{rb}|^2+\sigma_b^2}.
\end{align}
Again, in order avoid any impact of covert transmission on the transmission from Alice to Bob, we should guarantee $\gamma_b^0=\gamma_b^1$. Therefore, we have
\begin{align}
&\frac{P_r^0L_{rb}|h_{rb}|^2 G^2 P_a(1-\rho)L_{ar}|h_{ar}|^2}{P_r^0L_{rb}|h_{rb}|^2 G^2 \sigma^2_r+\sigma_b^2} \notag \\
&=\frac{P_r^1L_{rb}|h_{rb}|^2G^2P_a(1-\rho)L_{ar}|h_{ar}|^2}{P_r^1L_{rb}|h_{rb}|^2G^2\sigma^2_r+P_r^cL_{rb}|h_{rb}|^2+\sigma_b^2} .
\end{align}
After some algebra manipulations, the above equation can be rewritten as
\begin{align} \label{Pr1_PS}
P_r^1&= \frac{P_r^0\left(\eta_1\rho P_a L_{ar}|h_{ar}|^2L_{rb}|h_{rb}|^2+\sigma_b^2\right)}{P_r^0L_{rb}|h_{rb}|^2+\sigma_b^2}  \\
&=\frac{\eta_0\rho P_aL_{ar}|h_{ar}|^2\!\left(\eta_1\rho P_aL_{ar}|h_{ar}|^2L_{rb}|h_{rb}|^2\!+\!\sigma_b^2\right)}{\eta_0\rho P_aL_{ar}|h_{ar}|^2L_{rb}|h_{rb}|^2+\sigma_b^2}. \notag
\end{align}
As per \eqref{Pr0_PS} and \eqref{Pr1_PS}, we again have
\begin{align}
P_r^1&=\frac{\eta_1\rho P_aL_{ar}|h_{ar}|^2L_{rb}|h_{rb}|^2+\sigma_b^2}{\eta_0\rho P_aL_{ar}|h_{ar}|^2L_{rb}|h_{rb}|^2+\sigma_b^2}P_r^0  \notag \\
&>P_r^0,
\end{align}
which means that the relay requires more power to forward $\mathbf{x}_a$ when it transmits covert information to Bob.
Based on \eqref{Pr1_plus_prc_PS} and \eqref{Pr1_PS}, the transmit power of the covert information at the relay is given by
\begin{align} \label{Prc_PS}
P_r^c&=\eta_1\rho P_a L_{ar} |h_{ar}|^2-P_r^1  \notag \\
&= \frac{(\eta_1-\eta_0) \rho P_aL_{ar}|h_{ar}|^2\sigma_b^2}{\eta_0\rho P_aL_{ar}|h_{ar}|^2L_{rb}|h_{rb}|^2+\sigma_b^2}.
\end{align}

\subsection{Detection Performance and Optimal Detection Threshold at Alice}

In this subsection, we present the optimal detection strategy at Alice and her detection performance limits. Firstly, a decision rule for an arbitrary detection threshold is constructed based on a sufficient test statistic. Then, we derive the detection performance in terms of the false alarm and miss detection rates for any given detection threshold. Finally, we analytically obtain the optimal detection threshold that minimizes the detection error probability.

According to Appendix \ref{app:proof optimal detector}, the sufficient statistic $T$ in the PS scheme is given by
\begin{align}\label{t_a_PS}
T =\left\{
  \begin{array}{ll}
    \eta_0\rho P_a L_{ar}^2|h_{ar}|^4 + \sigma_a^2, &~~~~\Hnull, \\
    \eta_1\rho P_a L_{ar}^2|h_{ar}|^4 + \sigma_a^2,  &~~~~\Halt.
  \end{array}
\right.
\end{align}

The false alarm and miss detection rates at Alice for an arbitrary detection threshold $\tau$ in the following theorem, based on which we will tackle the optimization of $\tau$ in Theorem~\ref{theorem5}.

\begin{theorem}\label{theorem4}
In the PS scheme, the false alarm and miss detection rates at Alice for an arbitrary detection threshold $\tau$ are, respectively, derived as
\begin{align}
\alpha  &=\left\{
  \begin{array}{ll}
    1,  &~~~~~~\tau<\sigma_a^2, \\
    \exp\left\{-\frac{1}{\lambda_{ar}}\sqrt{\frac{\tau-\sigma_a^2}{\eta_0\rho P_a L_{ar}^2}}\right\},  &~~~~~~\tau>\sigma_a^2,
  \end{array} \label{PFA}
\right.  \\
\beta&=\left\{
  \begin{array}{ll}
    0,  &\tau<\sigma_a^2, \\
    1-\exp\left\{-\frac{1}{\lambda_{ar}}\sqrt{\frac{\tau-\sigma_a^2}{\eta_1\rho P_a dL_{ar}^2}}\right\},  &\tau>\sigma_a^2.
  \end{array} \label{PMD}
\right.
\end{align}
\end{theorem}

\begin{IEEEproof}
For a given $\tau$, following \eqref{t_a_PS}, the false alarm rate and miss detection rate are, respectively, given by
\begin{align}
\alpha &=\mathcal{P}\left[\eta_0\rho P_a L_{ar}^2|h_{ar}|^4 + \sigma_a^2\geq\tau\right] \notag  \\
&=\left\{\!
  \begin{array}{ll}
    1,  &\tau<\sigma_a^2, \\
    \mathcal{P}\left[|h_{ar}|^4 \geq \frac{\tau-\sigma_a^2}{\eta_0\rho P_aL_{ar}^2}\right],  &\tau>\sigma_a^2,
  \end{array} \label{PFA_2}
\right.\\
\beta &=\mathcal{P}\left[\eta_1\rho P_a L_{ar}^2|h_{ar}|^4 + \sigma_a^2<\tau\right]   \notag  \\
&=\left\{\!
  \begin{array}{ll}
    0,  &\tau<\sigma_a^2, \\
    \mathcal{P}\left[|h_{ar}|^4 < \frac{\tau-\sigma_a^2}{\eta_1\rho P_aL_{ar}^2}\right],  &\tau>\sigma_a^2.
  \end{array} \label{PMD_2}
\right.
\end{align}
Considering $F_{|h_{ar}|^4}(x) = 1 - \exp(-\sqrt{x}/\lambda_{ar})$, we achieve the desired results in \eqref{PFA} and \eqref{PMD} after some algebra manipulations, as per \eqref{PFA_2} and \eqref{PMD_2}.
\end{IEEEproof}

\begin{theorem}\label{theorem5}
The optimal threshold that minimizes $\xi$ in the PS scheme is derived as
\begin{align} \label{tau_ast}
\tau^{\ast}=\sigma_a^2+\left[\frac{\lambda_{ar}\sqrt{ \rho P_aL_{ar}^2\eta_0 \eta_1}}{2(\sqrt{\eta_1}-\sqrt{\eta_0})}\ln\left(\frac{\eta_1}{\eta_0}\right)\right]^2.
\end{align}
\end{theorem}

\begin{IEEEproof}
Following \eqref{PFA} and \eqref{PMD}, we have the detection error probability at Alice as
\begin{align} \label{xi}
\xi=\left\{
  \begin{array}{ll}
    1, & \tau \leq \sigma_a^2, \\
    1+\exp\left\{-\frac{\sqrt{\tau-\sigma_a^2}}{\lambda_{ar}\sqrt{\eta_0\rho P_aL_{ar}^2}}\right\}\\
    -\exp\Big\{-\frac{\sqrt{\tau-\sigma_a^2}}{\lambda_{ar}\sqrt{\eta_1\rho P_aL_{ar}^2}}\Big\},  &\tau>\sigma_a^2.
  \end{array}
\right.
\end{align}
Again, Alice will not set $\tau \leq \sigma_a^2$, since $\xi = 1$ is the worst case for Alice. Following \eqref{xi}, we derive the first derivative of $\xi$ with respect to $\tau$ for $\tau>\sigma_a^2$ as ${\partial\xi}/{\partial\tau} = \kappa_3(\tau)\kappa_4(\tau)$, where
\begin{align}
\kappa_3(\tau)&\!\triangleq\! \frac{\exp\left(-\frac{\sqrt{\tau-\sigma_a^2}}{\lambda_{ar}\sqrt{\eta_1\rho P_aL_{ar}^2}}\right)}{2\lambda_{ar}\sqrt{\eta_1\rho P_aL_{ar}^2(\tau-\sigma_a^2)}},  \label{kappa1_tau}  \\
\kappa_4(\tau)&\!\triangleq \! 1\!-\!\sqrt{\frac{\eta_1}{\eta_0}}\exp\!\left[-\frac{\sqrt{\tau-\sigma_a^2}}{\lambda_{ar}\sqrt{ \rho P_aL_{ar}^2}}\left(\frac{1}{\sqrt{\eta_0}}\!-\!\frac{1}{\sqrt{\eta_1}}\right)\!\right]. \label{kappa2_tau}
\end{align}
We note that $\kappa_3(\tau)>0$ due to $\tau>\sigma_a^2$. Hence, the value of $\tau$ that ensures ${\partial \xi}/{\partial\tau}=0$ is the one that guarantees $\kappa_4(\tau) = 0$, which is given by
\begin{align} \label{tau_dag_definition}
\tau^{\dag}=\sigma_a^2+\left[\frac{\lambda_{ar}\sqrt{ \rho P_aL_{ar}^2\eta_0\eta_1}}{2(\sqrt{\eta_1}-\sqrt{\eta_0})}\ln\left(\frac{\eta_1}{\eta_0}\right)\right]^2.
\end{align}
We note that ${\partial \xi}/{\partial\tau}<0$ for $\tau<\tau^{\dag}$ and ${\partial \xi}/{\partial\tau}>0$ for $\tau>\tau^{\dag}$. This is due to the fact that the term $\kappa_4(\tau)$ given in \eqref{kappa2_tau} is monotonically increasing with respect to $\tau$. Noting that $\tau^{\dag}>\sigma_a^2$, we can conclude that $\tau^{\dag}$ minimizes $\xi$ for $\tau>\sigma_a^2$. Noting $\xi$ is a continuous function of $\tau$, we obtain the optimal detection threshold as $\tau^{\ast}$ given in \eqref{tau_ast}.
\end{IEEEproof}

\begin{corollary}\label{corollary3}
The minimum value of the detection error probability $\xi$ at Alice is derived as
\begin{align} \label{xi_ast_PS}
\xi^{\ast}=1-\varphi^{\frac{1}{2(1-\sqrt{\varphi})}}\left(\frac{1}{\sqrt{\varphi}}-1\right),
\end{align}
where we recall that $\varphi\triangleq \eta_0/\eta_1$.
\end{corollary}

\begin{IEEEproof}
Substituting $\tau^{\ast}$ into \eqref{xi}, we obtain the minimum value of $\xi$, which is given by
\begin{align} \label{xi_ast_raw_PS}
\xi^{\ast}&=1+\exp\left\{-\frac{\sqrt{\eta_1}}{2(\sqrt{\eta_1}-\sqrt{\eta_0})}\ln\left(\frac{\eta_1}{\eta_0}\right)\right\}- \notag \\
&~~~~\exp\left\{-\frac{\sqrt{\eta_0}}{2(\sqrt{\eta_1}-\sqrt{\eta_0})}\ln\left(\frac{\eta_1}{\eta_0}\right)\right\}.
\end{align}
Following \eqref{xi_ast_raw_PS}, we achieve the desired result in \eqref{xi_ast_PS} after some algebra manipulations.
\end{IEEEproof}

\begin{remark}\label{remark5}
Following Corollary~\ref{corollary1} and Corollary~\ref{corollary3}, we note that the minimum detection error probability at Alice is the same in the TS and PS schemes, which is unexpected. This means that, although the transmission strategies at the relay are different in the TS and PS schemes, the monitoring ability of Alice on the relay is the same, which only depends on $\varphi\triangleq \eta_0/\eta_1$. Finally, this leads to that our discussions and conclusions given in Remark~\ref{remark1}, Remark~\ref{remark2}, and Corollary~\ref{corollary2} on the TS scheme are also valid for the PS scheme.
\end{remark}

\subsection{Optimization of Covert Transmission}

After subtracting the corresponding component related to $\mathbf{x}_a$ from the received signal $\mathbf{y}_b$ given in \eqref{y_b_H1}, the effective received signal used to decode the covert information $\mathbf{x}_c$ at Bob is
\begin{align} \label{tilde y_b_H1_PS}
\tilde{\mathbf{y}}_b[i]&\!=\!\sqrt{\!P_r^cL_{rb}}h_{rb}\mathbf{x}_c[i]\!+\!\sqrt{\!P_r^1L_{rb}}h_{rb}G\Big(\!\sqrt{(1\!-\!\rho)}\hat{\mathbf{n}}_{r,a}[i]+ \notag \\
&~~~\hat{\mathbf{n}}_{r,c}[i]\Big)+\mathbf{n}_{b,a}[i]+\mathbf{n}_{b,c}[i].
\end{align}
Then, following \eqref{tilde y_b_H1_PS} the SNR for $\mathbf{x}_c$ is given by
\begin{align} \label{gamma_c_PS}
\gamma_c &= \frac{P_r^cL_{rb}|h_{rb}|^2}{P_r^1L_{rb}|h_{rb}|^2 G^2 \sigma^2_r +\sigma^2_b} \notag \\
&\overset{c}{=}\frac{(Q_4-Q_3)\sigma_b^2}{\frac{Q_3(Q_4+\sigma_b^2)\sigma_r^2}{(1-\rho)P_aL_{ar}|h_{ar}|^2+\sigma_r^2}+ (Q_3+\sigma_b^2)\sigma_b^2 },
\end{align}
where
\begin{align}
Q_3 &\triangleq  \eta_0\rho P_a L_{ar}|h_{ar}|^2L_{rb}|h_{rb}|^2,    \notag \\
Q_4 &\triangleq   \eta_1\rho P_a L_{ar}|h_{ar}|^2L_{rb}|h_{rb}|^2,
\end{align}
and $\overset{c}{=}$ is obtained based on \eqref{G_PS}, \eqref{Pr1_PS}, and \eqref{Prc_PS}.
 As per \eqref{gamma_c_PS}, considering Rayleigh fading for $h_{ar}$ and $h_{rb}$, the average rate of the covert transmission from the relay to Bob in the PS scheme is given by
\begin{align} \label{rate_PS}
&C=\int_0^{\infty}\int_0^{\infty}\log_2\left\{1+\gamma_c\right\}f_{|h_{ar}|^2}(x)f_{|h_{rb}|^2}(y)\mathrm{d}x \mathrm{d}y \notag \\
&=\frac{1}{\lambda_{ar}\lambda_{rb}}\int_0^{\infty}\int_0^{\infty}\exp{\left[-\left(\frac{x}{\lambda_{ar}}+\frac{y}{\lambda_{rb}}\right)\right]}\times \notag \\
&~\log\!\left\{1\!+\!\frac{[Q_4(x,y)-Q_3(x,y)]\sigma_b^2}{\frac{Q_3(x,y)[Q_4(x,y)+\sigma_b^2]\sigma_r^2}{(1-\rho)P_aL_{ar}x+\sigma_r^2}\!+ \![Q_3(x,y)+\sigma_b^2] \sigma_b^2}\!\right\}\mathrm{d}x\mathrm{d}y.
\end{align}
Since $T/2$ is the effective communication time between relay and Bob in one block, the effective covert rate in the PS scheme is defined as
\begin{align} \label{thoughput_PS}
\Psi&=\frac{(T/2)}{T}C \notag \\
&=\frac{C}{2}.
\end{align}
Then, the optimization problem at relay of maximizing the effective covert subject to a certain covert communication constraint is given by
\begin{equation}\label{P1_PS}
\begin{aligned}
\quad \underset{\eta_0 \leq \eta_1 \leq \eta_u}{\max} \quad &\Psi \\
\text{s. t.} \quad  &\xi^{\ast}(\varphi) \geq 1 -\epsilon.
\end{aligned}
\end{equation}
The maximum value of $\Psi$ is then achieved by substituting the optimal value of $\eta_1$ (which is derived in the following theorem) into \eqref{thoughput_PS}, which is denoted by $\Psi^{\ast}$.

\begin{theorem} \label{theorem6}
For a given conversion efficiency factor $\eta_0$ under $\Hnull$ at relay, the optimal value (i.e., minimum value) of $\eta_1$ that achieves the maximum effective covert rate $\Psi$ subject to the covert communication constraint $\xi^{\ast}(\varphi) \geq 1 - \epsilon$ is given by
\begin{align} \label{eta1_ast_PS}
\eta_1^{\ast}=\left\{
  \begin{array}{ll}
    \frac{\eta_0}{\varphi_{\epsilon}},  &\epsilon\leq \left(\frac{\eta_0}{\eta_u}\right)^{\frac{\sqrt{\eta_u}}{2(\sqrt{\eta_u}-\sqrt{\eta_0})}}\left(\sqrt{\frac{\eta_u}{\eta_0}}-1\right), \\
    \eta_u,  &\mathrm{otherwise},
  \end{array}
\right.
\end{align}
where we recall that $\varphi_{\epsilon}$ is the solution of $\varphi$ to $\xi^{\ast}(\varphi)=1-\epsilon$.
\end{theorem}

\begin{IEEEproof}
The proof of Theorem~\ref{theorem6} is omitted here, which is similar to that of Theorem~\ref{theorem3}.
\end{IEEEproof}

\begin{remark}\label{remark6}
Based on Theorem~\ref{theorem3} and Theorem~\ref{theorem6}, we note that the optimal values of $\eta_1$ in the TS and PS schemes are the same for a predetermined $\eta_0$. This indicates that the constraint $\xi^{\ast} \geq 1 - \epsilon$ determines the same cost of achieving the covert transmission limits from the relay to Bob, in terms of the increase in the conversion efficiency factor, although the achieved maximum effective covert rates in the TS and PS schemes can be different.
\end{remark}

\section{Numerical Results}

In this section, we provide a thorough performance comparison between the TS and PS schemes. Based on our examination, we draw many useful insights with regard to the impact of some system parameters (e.g., $P_a$, $\eta_0$, and $\epsilon$ ) on covert transmission with harvested energy. Without other statements, we set $\lambda_{ar}=\lambda_{rb}=1$, $d_{ar}=d_{rb}=10$~m, $\sigma^2_{r,a}=\sigma^2_{r,c}=\sigma^2_{b,a}=\sigma^2_{b,c}=-80$~dBm, and $\eta_u=0.8$, the path loss exponent $m$ is set to 2, and carrier frequency $f_c$ is set to 900~MHz\cite{He2015Harvest}.

\begin{figure}
    \begin{center}
    \includegraphics[width=3.5in, height=2.9in]{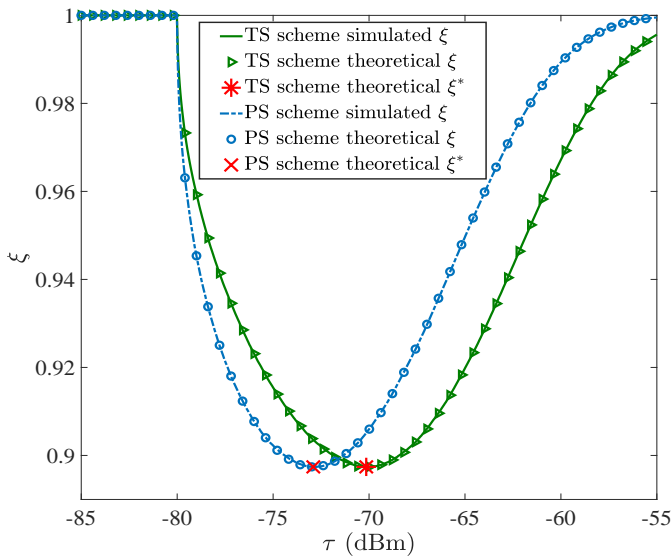}
    \caption{The detection error probability $\xi$ versus $\tau$, where $\eta_0=0.4$, $\eta_1=0.7$, and $P_a=20$~dBm.}\label{fig2}
    \end{center}
\end{figure}

In Fig.~\ref{fig2}, we plot the detection error $\xi$ versus Alice's detection threshold $\tau$ for the TS and PS schemes. As expected, we first observe that the simulated curves precisely match the theoretical ones, which confirms the correctness of our Theorem~\ref{theorem1} and Theorem~\ref{theorem4}. We also observe the minimum values of $\xi$ are equal in the TS and PS schemes, which verifies the correctness of our Theorem~\ref{theorem3} and Theorem~\ref{theorem6}, although the optimal detection thresholds that achieve these minimum values are different. In this figure, we further observe that the achieved detection error probability significantly varies with respect to the detection threshold, which demonstrates the importance of optimizing the detection threshold at Alice.

\begin{figure}
    \begin{center}
    \includegraphics[width=3.5in, height=2.9in]{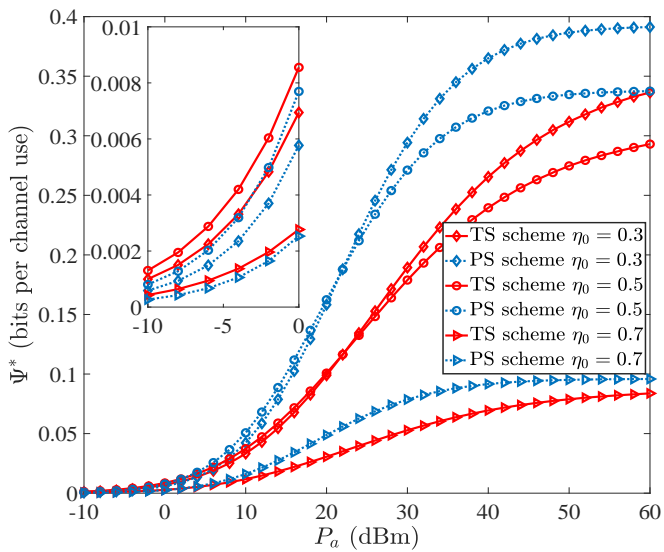}
    \caption{The effective covert rate $\Psi$ versus Alice's transmit power $P_a$ with different values of $\eta_0$, where $\epsilon=0.1$.}\label{fig3}
    \end{center}
\end{figure}

In Fig.~\ref{fig3}, we plot the maximum effective covert rate $\Psi^{\ast}$ versus $P_a$ with different values of $\eta_0$ for the TS and PS schemes. In this figure, we first observe that $\Psi^{\ast}$ monotonically increases as $P_a$ increases, which demonstrates that more covert information can be transmitted when more power is available at Alice and can be harvested at relay. In addition, in this figure we observe that the PS scheme outperforms the TS scheme when $P_a$ is in the high regime, since in this regime the transmit power is not the limited resource at the relay. However, when $P_a$ is smaller than some specific values (e.g., when $P_a \leq 0$~dBm), the performance of the TS scheme can be better than that of the PS scheme. This observation demonstrates the necessity of allowing the relay to switch between the TS and PS schemes (depending on the specific system parameters) in order to achieve a higher effective covert rate, which is the main motivation to propose these two schemes in this work.

\begin{figure}
    \begin{center}
    \includegraphics[width=3.5in, height=2.9in]{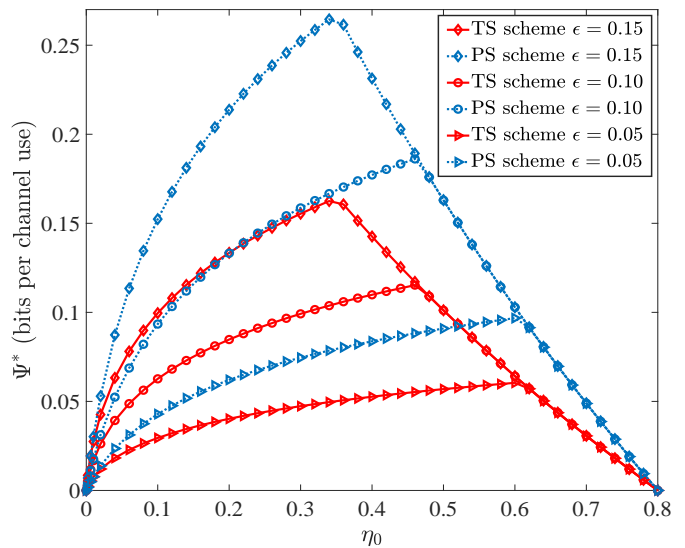}
    \caption{The maximum effective covert rate $\Psi^{\ast}$ versus $\eta_0$ with different values of $\epsilon$, where $P_a=20$~dBm.}\label{fig4}
    \end{center}
\end{figure}

In Fig.~\ref{fig4}, we plot the maximum effective covert rate $\Psi^{\ast}$ versus $\eta_0$ with different values of $\epsilon$.
In this figure, we first observe that $\Psi^{\ast} \rightarrow 0$ when $\eta_0 \rightarrow 0$, which is due to the fact that as $\eta_0 \rightarrow 0$ the relay cannot forward Alice's information to Bob and the covert transmission from the relay to Bob cannot be performed without the shield of its forwarding action. We also observe that $\Psi^{\ast} \rightarrow 0$ when $\eta_0 \rightarrow \eta_u$. This can be explained by the fact that as $\eta_0 \rightarrow \eta_u$ the relay cannot harvest extra energy from Alice to support its covert transmission. In addition, in this figure we observe that there is a sharp turning point on each curve of $\Psi^{\ast}$ versus $\eta_0$, which varies with the value of $\epsilon$. We confirm that this turning point occurs when ${\eta_0}/\varphi_{\epsilon}=\eta_u$, which can be explained by our Theorem~\ref{theorem3} and Theorem~\ref{theorem6}. This confirms that it is the covert communication constraint that limits $\Psi^{\ast}$ before the turning point, which can explain the observation that $\Psi^{\ast}$ increases with $\epsilon$ before all the turning points for each scheme. This also confirms that it is the energy harvester that limits $\Psi^{\ast}$ after each turning point, which can explain the observation that for different values of $\epsilon$ we may have the same $\Psi^{\ast}$ after all the turning points for each scheme.

\begin{figure}[!t]
    \begin{center}
    \includegraphics[width=3.5in, height=2.9in]{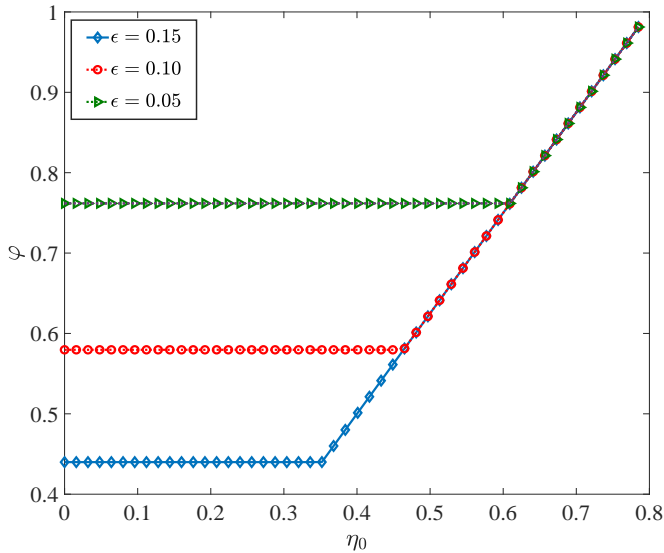}
    \caption{System overhead $\varphi$ versus $\eta_0$ under different values of $\epsilon$.}\label{fig5}
    \end{center}
\end{figure}

With the same system settings of Fig.~\ref{fig4}, in Fig.~\ref{fig5} we plot system overhead $\varphi$ (i.e., $\eta_0/\eta_1$) versus $\eta_0$ under different values of $\epsilon$, where $\eta_1$ is optimized by Theorem~\ref{theorem3} and Theorem~\ref{theorem6}. In this figure, we first observe that the values of $\varphi$ are consistent with $\varphi_{\epsilon}$ before $\eta_0$ reaches turning points on the curves, where $\varphi_{\epsilon}$ is the solution of $\varphi$ to $\xi^{\ast}(\varphi)=1-\epsilon$ and is solely determined by the given $\epsilon$. The value of the horizontal axis corresponding to turning points is denoted by $\eta_0^{\dag}$. When $\eta_0 \geq \eta_0^{\dag}$, we observe that $\eta_1^{\ast}$ is a monotonically increasing function of $\eta_0$.
Furthermore, we observe that the value of $\varphi_{\epsilon}$ decreases with $\epsilon$, which can be explained by our Remark~\ref{remark2} that $\xi^{\ast}$ is a monotonically increasing function of $\varphi$. It is illustrated that $\varphi \rightarrow 1$ when $\eta_0 \rightarrow \eta_u$, thus means that there is no enough opportunity for relay to harvest extra energy through a higher conversion efficiency factor $\eta_1$ when the value of $\eta_0$ is sufficiently large.

\begin{figure}
    \begin{center}
    \includegraphics[width=3.5in, height=2.9in]{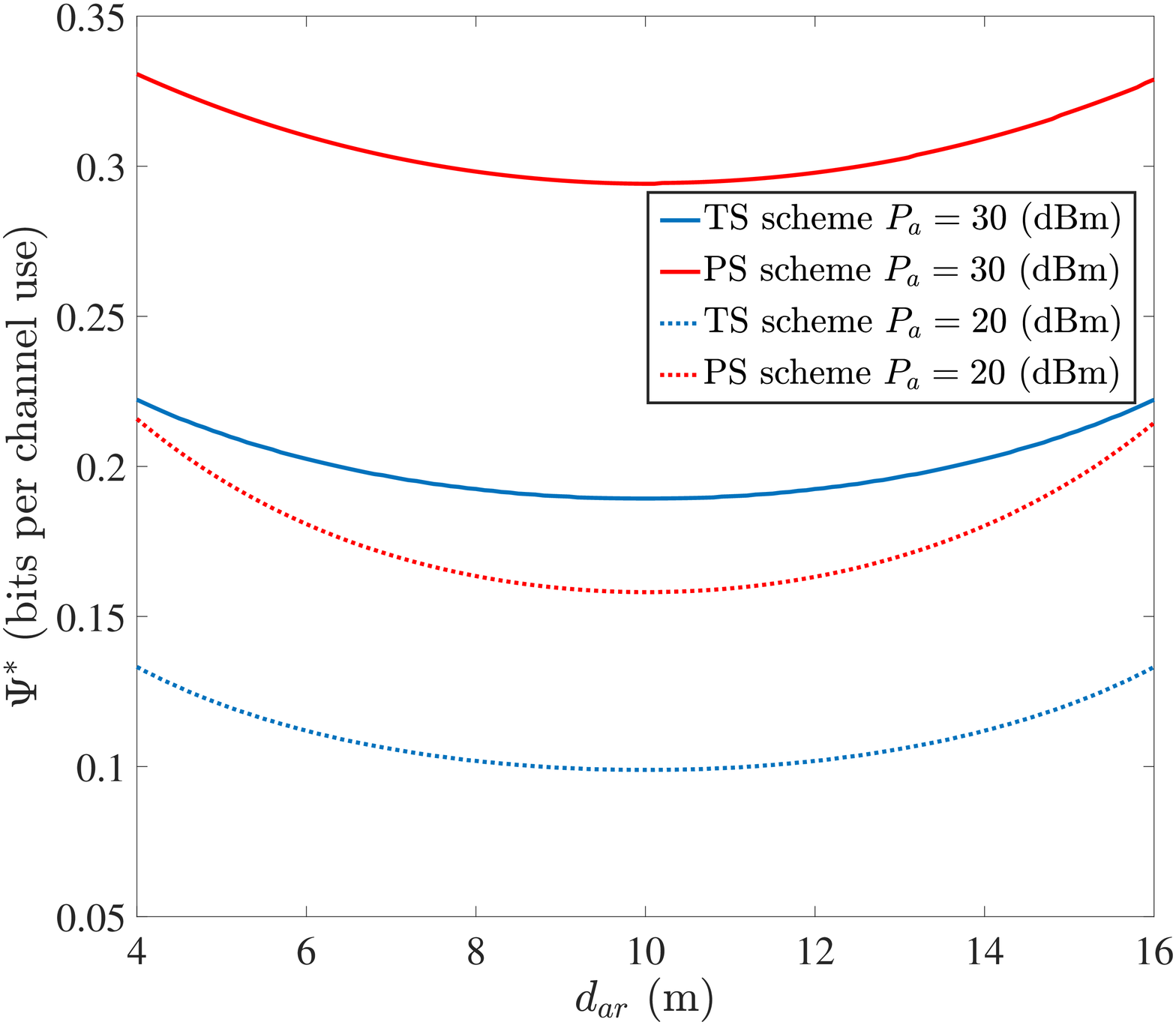}
    \caption{The maximum effective covert rate $\Psi^{\ast}$ versus the distance from Alice to the relay $d_{ar}$ with different values of $P_a$, where $d_{ar}+d_{rb}=20$~m, $\epsilon=0.1$.}\label{fig6}
    \end{center}
\end{figure}

In Fig.~\ref{fig6}, we plot the maximum effective covert rate $\Psi^{\ast}$ versus the distance from Alice to the relay $d_{ar}$ with different values of Alice's transmit power $P_a$. In this figure, we first observe that $\Psi^{\ast}$ first decreases and then increases as $d_{ar}$ increases, which indicates that there is a value of $d_{ar}$ (denoted by $d_{ar}^{\dag}$) that minimizes $\Psi^{\ast}$. When $d_{ar}$ is smaller than $d_{ar}^{\dag}$, $\Psi^{\ast}$ is monotonically decreasing with respect to $d_{ar}$, this can be explained by the fact that decreasing $d_{ar}$ simultaneously decreases the detection error probability at Alice but increases the power harvested by relay which is used for transmitting covert information to Bob, which means that $d_{ar}$ has a two-side impact on the considered covert communications.

\section{Conclusion}

This work examined the possibility, performance limits, and associated costs of covert communication achieved by a self-sustained relay over quasi-static Rayleigh fading channels, in which the relay opportunistically transmits its own information to the destination Bob covertly on top of forwarding Alice's information, while Alice tries to detect this covert transmission. Specifically, we considered the TS and PS schemes at the self-sustained relay for energy harvesting and analyzed Alice's detection performance limit in terms of the minimum detection error probability, based on which we determined the maximum effective covert rate $\Psi^{\ast}$ achieved subject $\xi^{\ast} \geq 1 - \epsilon$. Our analysis indicates that the required minimum energy conversion efficiency under $\Halt$, i.e., $\eta_1^{\ast}$, to achieve this $\Psi^{\ast}$ is the same for the TS and PS schemes, which indicates that the cost of achieving the relay's covert communication limits is the same, although the achievable $\Psi^{\ast}$ can be different. Our analysis also demonstrates that it is the constraint $\xi^{\ast} \geq 1 - \epsilon$ that limits $\Psi^{\ast}$ when $\epsilon$ is less than a specific value determined solely by $\eta_0$ and $\eta_u$, and otherwise it is $\eta_u$ that limits $\Psi^{\ast}$.

\appendices
\section{Proof of optimal detector} \label{app:proof optimal detector}
The optimality of radiometer can be proved along the same lines as the proof of Lemma 3 in \cite{Sobers2017Covert} using Fisher-Neyman factorization theorem and Likelihood Ratio Ordering concepts.

As per \eqref{fw_signal_TS} and \eqref{ya}, we note that the distribution of $h_{ra}$ is known to Alice, while its value in a given fading block time is not known, due to the fact that $\mathbf{x}_a$ has been modified to $\hat{\mathbf{x}}_a$ and channel estimation for $h_{ra}$ by using $\mathbf{x}_a$ is invalid. $\mathbf{y}_a[i]$ has a distribution given by
\begin{align}
\mathbf{y}_a[i]\!\sim\!\left\{
  \begin{array}{ll}
    \mathcal{CN}\left(0, P_r^0 L_{ra}|h_{ra}|^2\! + \!\sigma_a^2\right), &\Hnull, \\
    \mathcal{CN}\left(0, P_r^1 L_{ra}|h_{ra}|^2 \! + \! P_r^c L_{ra}|h_{ra}|^2\!+\!\sigma_a^2\right),  &\Halt,
  \end{array}
\right.
\end{align}
which can be rewritten as
\begin{align}
\mathbf{y}_a[i] \sim  \mathcal{CN}\left(0, \sigma_a^2+\theta\right),
\end{align}
where $\Hnull$ and $\Halt$ can be distinguished for $\Theta_0$ and $\Theta_1$ with probability distribution functions (PDFs) given by
\begin{align}
f_{\Theta_{\rho}}(\theta)=\left\{
  \begin{array}{ll}
    \frac{1}{P_r^0 L_{ra}|h_{ra}|^2}\times \\
    \mathrm{exp}\left(-\frac{\theta}{P_r^0 L_{ra}|h_{ra}|^2}\right), & 0<\theta, \rho=0  \\
    \frac{1}{(P_r^1+P_r^c) L_{ra}|h_{ra}|^2} \times \\
    \mathrm{exp}\left(-\frac{\theta}{(P_r^1+P_r^c) L_{ra}|h_{ra}|^2}\right),  & 0<\theta, \rho=1  \\
    0,  & \mathrm{Otherwise}.
  \end{array}
\right.
\end{align}
Here,
\begin{align}
\frac{f_{\Theta_1}(\theta)}{f_{\Theta_0}(\theta)}&=\frac{\frac{1}{(P_r^1+P_r^c) L_{ra}|h_{ra}|^2}\mathrm{exp}\left(-\frac{\theta}{(P_r^1+P_r^c) L_{ra}|h_{ra}|^2}\right)}{\frac{1}{P_r^0 L_{ra}|h_{ra}|^2}\mathrm{exp}\left(-\frac{\theta}{P_r^0 L_{ra}|h_{ra}|^2}\right)} \notag \\
&=\frac{P_r^0}{P_r^1+P_r^c}\mathrm{exp}\left(\frac{(P_r^1+P_r^c-P_r^0)\theta}{(P_r^1+P_r^c) P_r^0 L_{ra}|h_{ra}|^2}\right),
\end{align}
which is non-decreasing over the union of support of $\Theta_0$ and $\Theta_1$ due to the condition $P_r^1+P_r^c>P_r^0$ should be guaranteed in the considered scenario, thus $\Theta_0\leq_{\mathrm{lr}}\Theta_1$.

The distribution of Alice's observations conditioned over $\theta$ is
\begin{align}
f_{\mathbf{y}_a}(\theta)=\left(\frac{1}{\pi(\sigma_a^2+\theta)}\right)^n\mathrm{exp}\left[-\frac{\sum_{i=1}^n|\mathbf{y}_a[i]|^2}{\sigma_a^2+\theta}\right].
\end{align}
and according to the Fisher-Neyman factorization theorem, the total received power at Alice, $\sum_{i=1}^n|\mathbf{y}_a[i]|^2$ , is a sufficient statistic for Alice's test. The optimal decision rule for Alice is given by
\begin{align}\label{decisions}
\Lambda(\mathbf{y}_a)=\frac{\mathbb{E}_{\Theta_1}[f_{\mathbf{y}_a}(\theta)]}{\mathbb{E}_{\Theta_0}[f_{\mathbf{y}_a}(\theta)]}\mathop{\gtrless}\limits_{\Honull}^{\Hoalt}\Upsilon.
\end{align}
For $\mathbb{S}_{\mathbf{y}_a}\triangleq\sum_{i=1}^n|\mathbf{y}_a[i]|^2$, $\mathbb{S}_{\mathbf{y}_a}$ has a chi-squared distribution and from the definition of a chi-squared random variable,  $\mathbb{S}_{\mathbf{y}_a}(\theta)\leq_{\mathrm{lr}}\mathbb{S}_{\mathbf{y}_a}(\theta')$ for $\theta \leq \theta'$. Then the monotonicity of $\Lambda(\mathbf{y}_a)$ then follows from Stochastic ordering, and hence the likelihood ratio test is equivalent to
a threshold test on the received power.

While adopting a radiometer, the total received power at Alice, $\sum_{i=1}^n|\mathbf{y}_a[i]|^2$ is a sufficient statistic for Alice's test.
Since any one-to-one transformation of a sufficient statistic is also sufficient, the term $1/n\sum_{i=1}^n|\mathbf{y}_a[i]|^2$ is also a sufficient
statistic. Considering the infinite blocklength, i.e., $n \rightarrow \infty$, we have
\begin{align}\label{TT}
T &=\lim_{n \rightarrow \infty}\frac{1}{n}\sum_{i=1}^n|\mathbf{y}_a[i]|^2 \notag \\
&=\left\{
  \begin{array}{ll}
    P_r^0 L_{ra}|h_{ra}|^2+ \sigma_a^2, &~~~\Hnull, \\
    P_r^1 L_{ra}|h_{ra}|^2+ P_r^c L_{ra}|h_{ra}|^2+\sigma_a^2,  &~~~\Halt.
  \end{array}
\right.
\end{align}
Then, the decision rule in the adopted detector at Alice can be written as
\begin{align}\label{decisions}
T\mathop{\gtrless}\limits_{\Honull}^{\Hoalt}\tau,
\end{align}
where $\tau$ is the threshold for $T$, which will be optimized later in order to minimize the detection error probability. Therefore, $T$ is a sufficient test statistic and optimal detection threshold will be derived for the proposed schemes, thus makes the adopted radiometer be the optimal detector.

\bibliographystyle{IEEEtran}
\bibliography{IEEEfull,CC}

\begin{IEEEbiography}[{\includegraphics[width=1in,height=1.25in,clip,keepaspectratio]{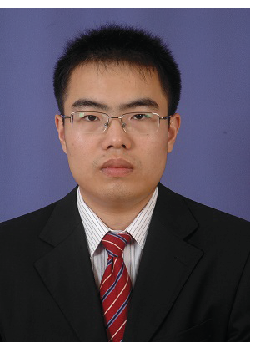}}]{Jinsong Hu} received the B.S. degree and Ph.D. degree from the School of Electronic and Optical Engineering, Nanjing University of Science and Technology, Nanjing, China in 2013 and 2018, respectively. From 2017 to 2018, he was a Visiting Ph.D. Student with the Research School of Engineering, Australian National University, Canberra, ACT, Australia. He is currently a Lecturer with the College of Physics and Information Engineering, Fuzhou University, Fuzhou, China. He served as a TPC member for the IEEE ICC 2019. His research interests include array signal processing, covert communications, and physical layer security.
\end{IEEEbiography}
\begin{IEEEbiography}[{\includegraphics[width=1in,height=1.25in,clip,keepaspectratio]{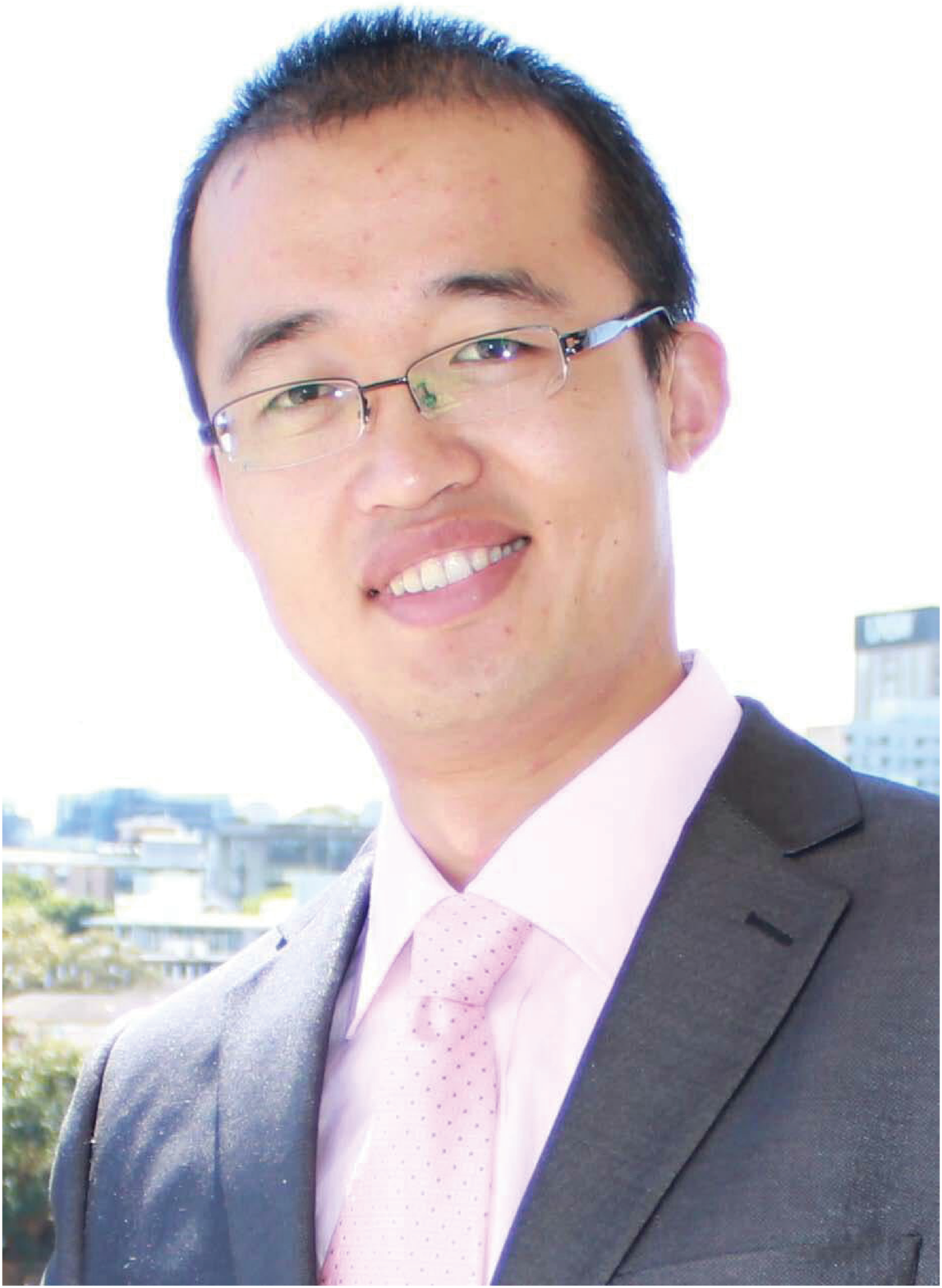}}]{Shihao Yan}
(S'11-M'15) received the Ph.D. degree in Electrical Engineering from The University of New South Wales, Sydney, Australia, in 2015. He received the B.S. in Communication Engineering and the M.S. in Communication and Information Systems from Shandong University, Jinan, China, in 2009 and 2012, respectively. From 2015 to 2017, he was a Postdoctoral Research Fellow in the Research School of Engineering, The Australian National University, Canberra, Australia. He is currently a University Research Fellow in the School of Engineering, Macquarie University, Sydney, Australia.
His current research interests are in the areas of wireless communications and statistical signal processing, including physical layer security, covert communications, and location spoofing detection.
\end{IEEEbiography}
\begin{IEEEbiography}[{\includegraphics[width=1in,height=1.25in,clip,keepaspectratio]{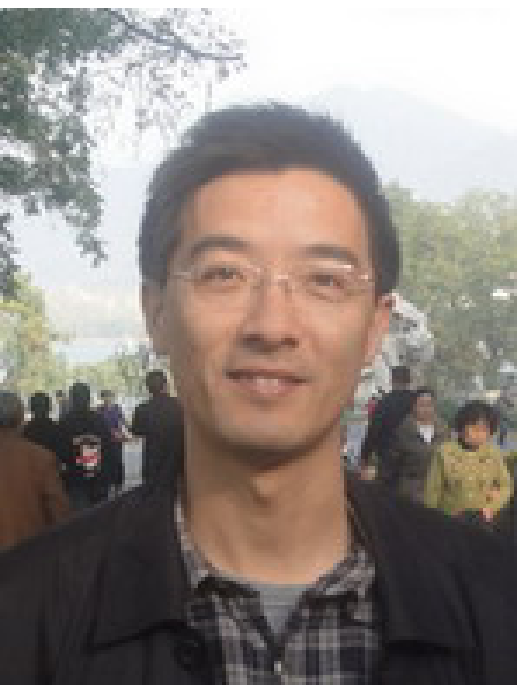}}]{Feng Shu}
was born in 1973. He received the B.S. degree from Fuyang Teachers College, Fuyang, China, in 1994, the M.S. degree from Xidian University, Xi'an, China, in 1997, and the Ph.D. degree from Southeast University, Nanjing, in 2002. In 2005, he joined the School of Electronic and Optical Engineering, Nanjing University of Science and Technology, Nanjing, China, where he is currently a Professor and a Supervisor of Ph.D. and graduate students. From 2009 to 2010, he held a visiting postdoctoral
position with The University of Texas at Dallas. He is also with Fujian Agriculture and Forestry University and awarded with Minjiang Scholar Chair Professor in Fujian Province. He has published
about 200 papers, of which over 100 are in archival journals, including over 40 papers on the IEEE journals and over 100 SCI-indexed papers. He holds ten Chinese patents. His research interests include wireless networks, wireless location, and array signal processing. He is currently an Editor of IEEE ACCESS.
\end{IEEEbiography}
\begin{IEEEbiography}[{\includegraphics[width=1in,height=1.25in,clip,keepaspectratio]{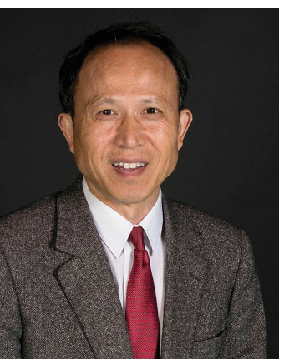}}]{Jiangzhou Wang}
(F'17) is currently a Professor and the former Head of the School of Engineering and Digital Arts at the University of Kent, U.K. He has published over 300 papers in international journals and conferences and four books in the areas of wireless mobile communications.

Professor Wang is a Fellow of the Royal Academy of Engineering, U.K., and a Fellow of the IEEE. He received the Best Paper Award from the IEEE GLOBECOM2012. He was an IEEE Distinguished Lecturer from 2013 to 2014. He was the Technical Program Chair of the 2019 IEEE International Conference on Communications (ICC2019), Shanghai, the Executive Chair of the IEEE ICC2015, London, and the Technical Program Chair of the IEEE WCNC2013. He has served as an Editor for a number of international journals, including IEEE Transactions on Communications from 1998 to 2013.
\end{IEEEbiography}

\end{document}